\documentclass[%
 aip,
 unsortedaddress,
 amsmath,amssymb,
reprint,%
]{revtex4-1}

\usepackage{hyperref}
\usepackage{physics}
\hypersetup{colorlinks=true, allcolors=blue}

\usepackage{graphicx}
\usepackage{dcolumn}
\usepackage{bm}
\usepackage{textcomp}
\usepackage{amsmath}
\def\mathclap#1{\text{\hbox to 0pt{\hss$\mathsurround=0pt#1$\hss}}}

\newcommand*{\citen}[1]{%
  \begingroup
    \romannumeral-`\x 
    \setcitestyle{numbers}%
    \citep{#1}%
  \endgroup   
}

\makeatletter
\newcommand{\vast}{\bBigg@{5}}
\newcommand{\Vast}{\bBigg@{5}}
\makeatother

\usepackage{xcolor}

\newcommand*{\colorboxed}{}
\def\colorboxed#1#{%
  \colorboxedAux{#1}%
}

\usepackage{tikz}
\usepackage{booktabs}
\usetikzlibrary{calc}

\usepackage{newfloat}
\DeclareFloatingEnvironment[
    fileext=los,
    listname={List of Schemes},
    name=MOV.,
    placement=tbhp,
]{movie}

\newcommand*{\colorboxedAux}[3]{%
  \begingroup
    \colorlet{cb@saved}{.}%
    \color#1{#2}%
    \boxed{%
      \color{cb@saved}%
      #3%
    }%
  \endgroup
}

\usepackage{subfigure}

\usepackage{multirow}
\usepackage{xcolor}
\usepackage{soul}

\usepackage[utf8]{inputenc}
\usepackage[T1]{fontenc}
\usepackage{mathptmx}
\usepackage{etoolbox}

\usepackage{lipsum}
\usepackage{mathtools}

\makeatletter
\def\@email#1#2{%
 \endgroup
 \patchcmd{\titleblock@produce}
  {\frontmatter@RRAPformat}
  {\frontmatter@RRAPformat{\produce@RRAP{*#1\href{mailto:#2}{#2}}}\frontmatter@RRAPformat}
  {}{}
}%
\makeatother
\begin{document}

\preprint{AIP/123-QED}

\title[Mn$_3$Sn APL Materials Draft]{Order parameter dynamics in Mn$_3$Sn driven by DC and pulsed spin-orbit torques}
\author{Ankit Shukla}
\author{Siyuan Qian}%
\author{Shaloo Rakheja}
 \email{ankits4@illinois.edu, siyuanq3@illinois.edu, rakheja@illinois.edu}
\affiliation{Holonyak Micro and Nanotechnology Laboratory, University of Illinois at Urbana-Champaign, Urbana, IL 61801}%

\date{\today}

\begin{abstract}
We numerically investigate and develop analytic models for both the DC and pulsed spin-orbit-torque (SOT)-driven response of order parameter in single-domain Mn$_3$Sn, which is a metallic antiferromagnet with an anti-chiral 120$^\circ$ spin structure. 
We show that DC currents above a critical threshold can excite oscillatory dynamics of the order parameter in the gigahertz to terahertz frequency spectrum. 
Detailed models of the oscillation frequency versus input current are developed and found to be in excellent agreement with the numerical simulations of the dynamics. 
In the case of pulsed excitation, the magnetization can be switched from one stable state to any of the other five stable states in the Kagome plane by tuning the duration or the amplitude of the current pulse. 
Precise functional forms of the final switched state versus the input current are derived, offering crucial insights into the switching dynamics of Mn$_3$Sn. 
The readout of the magnetic state can be carried out via either the anomalous Hall effect, or the recently demonstrated tunneling magnetoresistance in an all-Mn$_3$Sn junction. 
We also discuss possible disturbance of the magnetic order due to heating that may occur if the sample is subject to large currents. 
Operating the device in pulsed mode or using low DC currents reduces the peak temperature rise in the sample due to Joule heating.
Our predictive modeling and simulation results can be used by both theorists and experimentalists to explore the interplay of SOT and the order dynamics in Mn$_3$Sn, and to further benchmark the device performance.
\end{abstract}

\maketitle

\section{Introduction}
\vspace{-5pt}
Antiferromagnets (AFMs) are a class of magnetically ordered
materials that exhibit negligible net magnetization, owing to the unique arrangement of strongly exchange-coupled spins on the atoms of their unit cells. 
As a result, AFMs produce negligible stray fields, are robust to external magnetic field perturbations, and their precession frequency, set by the geometric mean of exchange and anisotropy energies, is in the terahertz (THz) regime.~\citep{gomonay2014spintronics, baltz2018antiferromagnetic, jungfleisch2018perspectives}   
The past decade has witnessed a rapid rise in theoretical and experimental research focused on the fundamental understanding and applications of AFM materials as active spintronic device elements.~\citep{jungwirth2016antiferromagnetic, cheng2016terahertz, khymyn2017antiferromagnetic, gomonay2018antiferromagnetic, jungfleisch2018perspectives, siddiqui2020metallic} 
There exists a broad range of AFM materials including insulators, metals, and semiconductors with unique properties that could be exploited to realize magnonic devices,~\citep{rezende2019introduction} high-frequency signal generators and detectors,~\citep{cheng2016terahertz, khymyn2017antiferromagnetic, sulymenko2017terahertz, gomonay2018narrow, parthasarathy2021precessional, zhao2021terahertz, shukla2022spin, zhao2022tunable} and non-volatile memory.~\citep{kosub2017purely, olejnik2018terahertz, han2023coherent}  
For example, insulators like NiO~\citep{hahn2014conduction, rezende2016diffusive} and MnF$_2$~\citep{wu2016antiferromagnetic} are well studied and have the potential to carry charge-less spin waves or magnons. Insulating AFM Cr$_2$O$_3$ shows magnetoelecticity below its N\'eel temperature of 307 K, which was exploited to demonstrate voltage-controlled exchange-bias memory and fully electrically controlled memory devices.~\citep{mahmood2021voltage}  
On the other hand, metallic AFMs have been mostly used as sources of exchange bias in spin valves and tunnel junction-based devices.~\citep{nogues1999exchange, zhang2016epitaxial} More recently, there has been significant research activity in non-collinear, chiral metallic AFMs of the form Mn$_3$X with a triangular spin structure and several intriguing magneto-transport characteristics such as a large spin Hall effect (SHE),~\citep{zhang2016giant} anomalous Hall effect (AHE),~\citep{kubler2014non, zhang2017strong, iwaki2020large} and ferromagnet-like spin-polarized currents.~\citep{vzelezny2017spin} 

Negative chirality materials, such as Mn$_3$Sn, Mn$_3$Ge, Mn$_3$Ga, perhaps best represent the promise of non-collinear metallic AFMs with a potential for ferromagnet-like spintronic devices in which the order parameter could be fully electrically controlled and read-out.~\citep{yan2022quantum, wang2022noncollinear, wang2023spin} 
In a recent experiment,~\citep{takeuchi2021chiral} conducted in a bilayer of heavy metal and Mn$_3$Sn, a characteristic fluctuation of the Hall resistance was measured in response to a DC current in the heavy metal. This observation could be explained in terms of the rotation of the chiral spin structure of Mn$_3$Sn driven by spin-orbit torque (SOT). 
Pal et al. and Krishnaswamy et al. independently showed that Mn$_3$Sn layers thicker than the spin diffusion length could be switched by seeded SOTs.~\citep{pal2022setting, krishnaswamy2022time} 
Here, the SOT sets the spin texture of the AFM in a thin layer at the interface, which acts as the seed for the subsequent setting of the domain configuration of the entire layer. The seeded SOT also requires bringing the temperature of the AFM above its ordering temperature and then cooling it in the presence of the SOT generated in a proximal heavy metal layer. 
Very recently, tunneling magnetoresistance (TMR) of approximately 2\% at room temperature in an all antiferromagnetic tunnel junction consisting of Mn$_3$Sn/MgO/Mn$_3$Sn was experimentally measured.~\citep{chen2023octupole} The TMR in Mn$_3$Sn originates from the time reversal symmetry breaking and the momentum-dependent spin splitting of bands in the crystal.
These recent works highlight the tremendous potential of Mn$_3$Sn and other negative chirality AFMs to explore and develop spintronic device concepts.

In this paper, we discuss the energy landscape of a thin film of Mn$_3$Sn in the mono-domain limit and deduce the weak six-fold magnetic anisotropy of the film via perturbation and numerical solutions (Section~\ref{sec:energy}). Consequences of the six-fold anisotropy on the equilibrium states, the origin of the weak ferromagnetic moment, and SOT-induced non-equilibrium dynamics are carefully modeled in Sections~\ref{sec:energy} and~\ref{sec:SOT}. The analytic model of the threshold spin current to drive the system into steady-state oscillations is validated against numerical simulations of the equation of motion. Because of the weak in-plane magnetic anisotropy, on the order of 100 J/m$^3$, we find that the critical spin current to induce dynamic instability of the order parameter in Mn$_3$Sn could be orders of magnitude lower than that in other AFMs such as NiO~\citep{cheng2016terahertz, khymyn2017antiferromagnetic, parthasarathy2021precessional} and Cr$_2$O$_3$~\cite{parthasarathy2021precessional}. 
Additionally, the oscillation frequency of the order parameter in Mn$_3$Sn could, in principle, be tuned from the gigahertz (GHz) to the terahertz (THz) scale~\citep{shukla2022spin}.
We also examine the response of the system when subject to pulsed spin current. Our results show that by carefully tuning the pulse width of the current ($t_\mathrm{pw}$) and the spin charge density injected into the system ($J_s t_\mathrm{pw}$, where $J_s$ is the spin current density) the evolution of the order parameter across different energy basins, separated by $60^\circ$ in the phase space, can be controlled. 
Our results of the SOT-driven dynamics in single-domain Mn$_3$Sn are closely related to the experimental results of Ref.~\citen{takeuchi2021chiral}.
Previous theoretical works on the current-driven dynamics in collinear as well as non-collinear AFMs have only addressed systems with zero net magnetization and those with two-fold anisotropy.~\citep{gomonay2012symmetry, gomonay2015using, shukla2022spin}. 
Another work focusing on NiO with six-fold anisotropy had only presented numerical aspects of pulsed SOT switching.~\citep{chirac2020ultrafast}
On the other hand, here, we address both the numerical and analytical aspect of current-driven dynamics in six-fold anisotropic system with a small non-zero net magnetization. 
We also discuss the AHE and TMR detection schemes and present the magnitude of output voltages for typical material parameters in both the cases (Section~\ref{sec:detect}).
Finally, we highlight the impact of thermal effects on the dynamics of the order parameter and the temperature rise of the sample when DC spin currents are acting on Mn$_3$Sn to generate high-frequency oscillations (Section~\ref{sec:temp}). Our models highlight the limits, potential, and opportunities of negative chirality metallic AFMs, prominently Mn$_3$Sn, for developing functional magnetic devices that are fully electrically controlled. 

\section{Crystal and Spin Structure}
\begin{figure}[ht!]
  \centering
  \includegraphics[width = \columnwidth, clip = true, trim = 0mm 0mm 0mm 0mm]{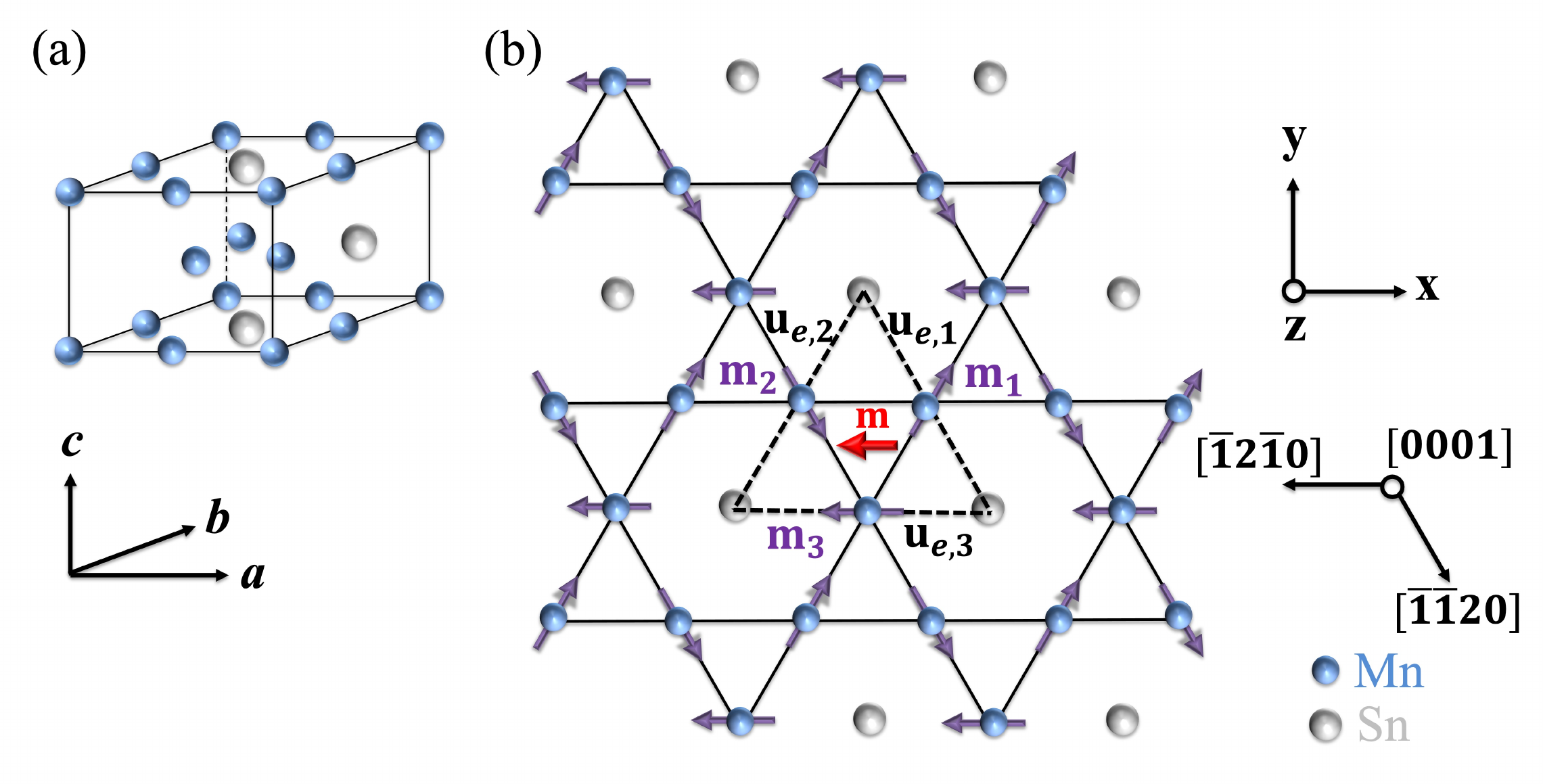}
  \caption{(a) Layered hexagonal crystal structure of Mn$_3$Sn.
  The $a$ and $c$ axes are parallel to the $[\overline{1}2\overline{1}0]$ and $[0001]$ directions, respectively.
  (b) Hexagonal crystal and inverse triangular spin structure of Mn$_3$Sn in the basal plane. In each hexagon the Mn atoms are at the corners (blue) while the Sn atoms are at the center (gray). 
  The spins (purple arrows) on neighboring Mn atoms are aligned at an angle of $120^\circ$ with respect to each other.
  The dashed lines connecting the Sn atoms represent the three easy axes ($\vb{u}_{e, i}$) corresponding to the magnetocrystalline anisotropy. 
  They are ordered counterclockwise whereas $\vb{m}_1, \vb{m}_2$, and $\vb{m}_3$ are ordered clockwise.
  The red arrow represents the net magnetization.
  The z-axis coincides with $[0001]$.} 
  \label{fig:crystal}
\end{figure}

\vspace{-5pt}
Mn$_3$Sn has a N\'eel temperature of approximately $420~\mathrm{K}$ and can only be stabilized in excess of Mn atoms.  
Below this temperature, it crystallizes into a layered hexagonal $D0_{19}$ structure, where the Mn atoms form a Kagome-type lattice in basal planes that are stacked along the $c$ axis ($[0001]$ direction), as shown in Fig.~\ref{fig:crystal}(a).
In each plane, the Mn atoms are located at the corners of the hexagons whereas the Sn atoms are located at their respective centers.
The magnetic moments on the Mn atoms, on the other hand, have been shown by neutron diffraction experiments~\citep{tomiyoshi1982magnetic} to form a noncollinear triangular spin structure with negative chirality, as shown in Fig.~\ref{fig:crystal}(b).
The nearest neighbor Mn moments ($\vb{m}_1, \vb{m_2}$ and $\vb{m}_3$) are aligned at an angle of approximately $120^\circ$, with respect to each other, resulting in a small net magnetic moment $\vb{m}$.
The weak ferromagnetism in Mn$_3$Sn, and similar antiferromagnets such as Mn$_3$Ge and Mn$_3$GaN, has been attributed to the geometrical frustration of the triangular antiferromagnetic structure, which leads to slight canting of Mn spins toward in-plane easy axes.~\citep{markou2018noncollinear}
The chirality of the spin structure breaks time reversal symmetry and leads to momentum-dependent Berry curvature and modifies the magneto-transport signals in Mn$_3$Sn.~\citep{kubler2014non, markou2018noncollinear} 
Switching from one chirality to another flips the sign of the Berry curvature and thus the sign of magneto-transport signals that are odd with respect to Berry curvature.  

Previous theoretical works~\citep{liu2017anomalous, li2022free} posit the aforementioned crystal structure to a hierarchy of interactions typical for 3d transition metal ions---strong Heisenberg exchange, followed by weaker Dzyaloshinskii-Moriya (DM) interaction, with single-ion anisotropy being the weakest.
Their model of the magnetic free energy interaction in a unit cell of bulk Mn$_3$Sn crystal, based on this hierarchy of interactions, predicts the existence of a small in-plane net magnetization, six-fold anisotropy degeneracy, and the deformation of the magnetic texture by an external magnetic field.
The basic spin structure of the free energy model in case of Mn$_3$Sn comprises two triangles, with three spins each, one on each basal plane.~\citep{liu2017anomalous, li2022free} 
While the free energy model of Ref.~\citen{liu2017anomalous} explicitly considers an interplane exchange interaction in addition to the intraplane exchange interaction, the free energy model of Ref.~\citen{li2022free} considers an effective exchange parameter, which includes both these effects.
The latter approach simplifies the system of six spins to that of three spins. 
Therefore, in our work, we assume a free energy model based on the latter approach viz., a system of three magnetic moments.~\citep{yamane2019dynamics, shukla2022spin}

\vspace{-10pt}
\section{Free Energy Model and Equilibrium States}\label{sec:energy}
\vspace{-5pt}
We consider a single domain non-collinear antiferromagnetic particle of Mn$_3$Sn of size $L_a \times w_a \times d_a$.
In the continuum modeling approach, the single domain Mn$_3$Sn is assumed to be composed of three equivalent interpenetrating sublattices, which are represented by magnetization vectors $\vb{m}_{1}, \vb{m}_{2}$, and $\vb{m}_{3}$.~\citep{takeuchi2021chiral, shukla2022spin}
Each magnetization vector has a constant saturation magnetization, $M_s$, and is strongly coupled to the other by a symmetric exchange interaction, characterized by the exchange constant $J_E (>0)$.
In addition, an asymmetric Dzyaloshinskii-Moriya interaction (DMI), characterized by the coefficient $D (>0)$, and single-ion uniaxial magnetocrystalline anisotropy, characterized by constant $K_{e} (>0)$ are assumed to describe the system. 
The free energy density is, therefore, defined as~\citep{yamane2019dynamics, shukla2022spin}
\begin{flalign}\label{eq:energy_density}
    \begin{split}
        F\qty(\vb{m}_1, \vb{m}_2, \vb{m}_3) &=  J_E\qty(\vb{m}_{1} \vdot \vb{m}_{2} + \vb{m}_{2} \vdot \vb{m}_{3} + \vb{m}_{3} \vdot \vb{m}_{1}) \\
        &+ D \vb{z} \vdot \left(\vb{m}_{1} \cp \vb{m}_{2} + \vb{m}_{2} \cp \vb{m}_{3} + \vb{m}_{3} \cp \vb{m}_{1} \right) \\
        & - \sum_{i = 1}^{3}\qty(K_{e} \qty(\vb{m}_{i} \vdot \vb{u}_{e, i})^{2} + \mu_{0}M_{s}\vb{H}_{a} \vdot \vb{m}_{i}),
    \end{split}
\end{flalign}
where $\vb{u}_{e, i}$ is the easy axis corresponding to $\vb{m}_i$. Here, they are assumed to be $\vb{u}_{e, 1} = -(1/2) \vb{x} + (\sqrt{3}/2) \vb{y}$, $\vb{u}_{e, 2} = -(1/2) \vb{x} - (\sqrt{3}/2) \vb{y}$ and $\vb{u}_{e, 3} = \vb{x}$, respectively.
Typical value of these energy constants are listed in Table~\ref{tab:mat_params1}.
The last term in Eq.~(\ref{eq:energy_density}) is the Zeeman energy due to the externally applied magnetic field $\vb{H}_a$. 
For the results presented in this work no external magnetic field is considered.

Compared to previous works discussing micromagnetic modeling of AFMs,~\citep{puliafito2019micromagnetic, shukla2022spin} monodomain modeling is applicable for AFM particles that are smaller than the domain size, which is typically $200-250~\mathrm{nm}$ for Mn$_3$Sn.~\citep{takeuchi2021chiral, higo2022perpendicular}
For such length scales spatial variation in exchange and DM interactions can be safely ignored.\citep{sato2023thermal}
The details of the micromagnetic simulation framework, including the boundary conditions, can be consulted from our previous work.~\citep{shukla2022spin}
Since in this work we focus on the mono-domain limit, the cell dimensions are the same as the physical dimensions of the AFM layer. The thickness of the AFM film plays a role in establishing the spin angular momentum acting on the AFM from the proximal NM.
In contrast to the continuum modeling approach presented in this work and in Ref.~\citen{takeuchi2021chiral}, some previous works have employed a three spin atomistic model.~\citep{tsai2020electrical, pal2022setting, higo2022perpendicular}
Although the atomistic modeling approach is more accurate, it becomes computationally expensive for large systems of size 10's of nm and above since the number of atoms and the associated spins increase.
On the other hand, the continuum modeling framework is more suitable for large systems and has been shown to agree well with atomistic simulations in the case of another manganese-based AFM, Mn$_2$Au.~\citep{hirst2022multiscale}
Therein, the material parameters used in the continuum framework were well calibrated against those in the atomistic model. 
The material parameters used in our work are also calibrated against an atomistic model as mentioned in Ref.~\citen{yamane2019dynamics}

In exchange energy dominant AFMs, such as Mn$_3$Sn, the hierarchy of energy interactions leads to $J_E \gg D \gg K_e$. 
As a result, the exchange energy is minimized if the three sublattice vectors are at an angle of $\frac{2\pi}{3}$ with respect to each other. The minimization of the DM interaction energy confines the sublattice vectors to the x-y plane with zero z-component, and enforces a clockwise ordering between the vectors $\vb{m}_1$, $\vb{m}_2$, and $\vb{m}_3$.
The magnetocrystalline anisotropy energy would be minimized if each $\vb{m}_i$ coincides with its respective $\vb{u}_{e, i}$. 
However, counterclockwise ordering of $\vb{u}_{e, 1}$, $\vb{u}_{e, 2}$, and $\vb{u}_{e, 3}$ along with the clockwise ordering of $\vb{m}_1$, $\vb{m}_2$, and $\vb{m}_3$ implies that all the magnetization vectors cannot coincide with their respective easy axis simultaneously. 
This would lead to six equilibrium or minimum energy states wherein only of the sublattice vectors is coincident with its easy axis.

\begin{figure*}[ht!]
  \centering
  \includegraphics[width = \textwidth, clip = true, trim = 0mm 0mm 0mm 0mm]{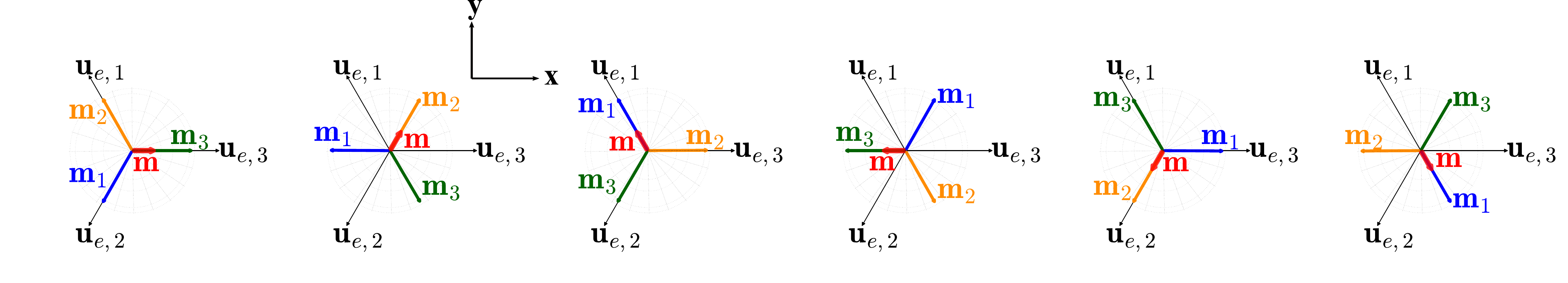}
  \caption{Six possible equilibrium states in single domain Mn$_3$Sn crystal. They lie in the Kagome plane, which is assumed to coincide with the x-y plane, and are separated from each other by $60$ degrees. 
  In each case, only one of the sublattice vectors, $\vb{m}_\mathrm{i}$, coincides with its easy axis, $\vb{u}_{e, \mathrm{i}}$. A small in-plane average magnetization, $\vb{m}$, also coincides with this particular $\vb{m}_i$ and its corresponding easy axis. Here, $\vb{m}$ is not drawn to scale but magnified by $100 \times$ for the purpose of clear representation.} 
  \label{fig:equilibrium}
\end{figure*}

\begin{table}[ht!]
\caption{\label{tab:mat_params1} {List of material parameters for the AFM, Mn$_3$Sn, and the heavy metal (HM), which is chosen as W, in the SOT device setup.}}
\begin{ruledtabular}
\begin{tabular}{lccr}
Parameters  & Definition  &  Values & Ref.\\
\hline
$J_E$ & Exchange constant &$2.4\times10^8~\mathrm{J/m^3}$ & \citen{yamane2019dynamics}\\
$D$  & DMI constant &$2\times10^7~\mathrm{J/m^3}$ & \citen{yamane2019dynamics}\\
$K_e$ & Uniaxial anisotropy constant &$3\times10^6~\mathrm{J/m^3}$  & \citen{yamane2019dynamics} \\
$M_s$ & Saturation magnetization &$1.63~\mathrm{T}$ &\citen{yamane2019dynamics}\\
$\alpha$ & Gilbert damping & $0.003$ & \citen{tsai2020electrical}\\
$d_\mathrm{HM}$ & Thickness of HM & $7~\mathrm{nm}$ & \citen{higo2022perpendicular}\\
$\rho_\mathrm{HM}$ & Resistivity of HM & $43.8~\mathrm{\mu \Omega \vdot cm}$ & \citen{higo2022perpendicular}\\
$\theta_\mathrm{SH}$ & Spin Hall angle for HM & $0.06$ & \citen{higo2022perpendicular}\\
\end{tabular}
\end{ruledtabular}
\end{table}
Minimizing Eq.~(\ref{eq:energy_density}) with respect to $\vb{m}_{i}$ leads to six equilibrium states, as shown in Fig.~\ref{fig:equilibrium}.
They are given as $\varphi_m^\mathrm{eq} = n\pi/3$, where $n = \{0, 1, 2, 3, 4, 5\}$.
In each case, only one of the three magnetization vectors coincides with its easy axis. 
A small in-plane average magnetization, $\vb{m} = \frac{\vb{m}_1 + \vb{m}_2 + \vb{m}_3}{3}$, is also obtained, depicted by the red arrows in Fig.~\ref{fig:equilibrium}. 
In each case, $\vb{m}$ coincides with the sublattice vector that is aligned along its easy axis. 
Our numerical calculations reveal the norm of the average magnetization to be equal but
very small for all the six ground states, approximately $\norm{\vb{m}} \approx 3.66 \times 10^{-3}$.
Therefore, in Fig.~\ref{fig:equilibrium}, $\vb{m}$ is zoomed-in by $100 \times$ for the sake of clear presentation.
The small non-zero value of $\norm{\vb{m}}$ suggests that the angle between the sublattice vectors is not exactly $120^\circ$, but deviates slightly.
Defining this deviation as $\eta_{ij} = \cos^{-1}{(\vb{m}_i \vdot \vb{m}_j)} - \frac{2\pi}{3}$, we find that $\eta_{ij} \approx -0.36^\circ$ if either of $\vb{m}_i$ or $\vb{m}_j$ is coincident with its easy axis while $\eta_{ij} \approx 0.72^\circ$ if neither $\vb{m}_i$ nor $\vb{m}_j$ coincides with its easy axes.
This small canting of sublattice vectors is well know, as mentioned in previous theoretical and experimental works.~\citep{tomiyoshi1982magnetic, nagamiya1982triangular, markou2018noncollinear}
Our numerical results show that $\eta_{12} + \eta_{23} + \eta_{31} = 0$.

To gain further insight into the six-fold degenerate equilibrium states, we adopt a perturbative approach~\citep{liu2017anomalous, li2022free, zhang2023current} to derive a simple model of the energy density.
Firstly, the order parameters, viz. sublattice vectors and the average magnetization vector, in the equilibrium states are assumed to be in the x-y plane. 
Secondly, the angle between $\vb{m}_i$ and $\vb{m}_j$ is assumed to deviate slightly from $120^\circ$. 
Therefore, we have
\begin{equation}\label{eq:avg_m}
    \vb{m} = u_m\begin{pmatrix}
                            \cos{\varphi_m} \\
                            \sin{\varphi_m} \\
                            0 \\
                            \end{pmatrix}, 
\end{equation}
and
\begin{equation}\label{eq:m_i}
    \vb{m}_i = \begin{pmatrix}
                              \cos{\qty(\phi_i)} \\
                              \sin{\qty(\phi_i)} \\
                              0
                \end{pmatrix} = \begin{pmatrix}
                              \cos{\qty(-\varphi_m - \frac{2\pi i}{3} + \eta_i)} \\
                              \sin{\qty(-\varphi_m - \frac{2\pi i}{3} + \eta_i)} \\
                              0
                \end{pmatrix},
\end{equation}
where $u_m = \norm{\vb{m}}$, $\varphi_m$ is the azimuthal angle corresponding to $\vb{m}$, $i = \{1, 2, 3\}$, while $\eta_i$ is the small deviation from the perfect $\frac{2\pi}{3}$ ordering $(i.e. \eta_i \ll \frac{2\pi}{3})$. 
They are defined such that $\eta_1 + \eta_2 + \eta_3 = 0$. 
Using the perturbative approach, also presented in the supplementary material, we find 
\begin{subequations}\label{eq:eta_i_1}
    \begin{align}
        \eta_1 &\approx \frac{K_e}{3\qty(J_E + \sqrt{3} D)} \qty(\sqrt{3} \cos{(2\varphi_m)} - \sin{(2\varphi_m)}), \\
        \eta_2 &\approx -\frac{K_e}{3\qty(J_E + \sqrt{3} D)} \qty(\sqrt{3} \cos{(2\varphi_m)} + \sin{(2\varphi_m)}),
    \end{align}
\end{subequations}
which is then used to obtain $u_m$ as
\begin{equation}\label{eq:um}
    u_m = \frac{K_e}{3\qty(J_E + \sqrt{3} D)}.
\end{equation}
On further analysis we obtain the energy density as  
\begin{flalign}\label{eq:energy_density_phim}
    \begin{split}
        F\qty(\varphi_m) &\approx  - \frac{\qty(3J_E + 7\sqrt{3}D)K_e^3}{18 \qty(J_E + \sqrt{3}D)^3} \cos{\qty(6\varphi_m)},
    \end{split}
\end{flalign}
where the constant energy terms are not shown. This $\cos{\qty(6\varphi_m)}$ dependence explains the six minimum energy states shown in Fig.~\ref{fig:equilibrium}. 
We find the analytic expressions of Eqs.~(\ref{eq:eta_i_1}) and~(\ref{eq:um}) to match very well against the numerical results (presented in supplementary material).

\begin{figure}[ht!]
  \centering
  \includegraphics[width = \columnwidth, clip = true, trim = 0mm 0mm 0mm 0mm]{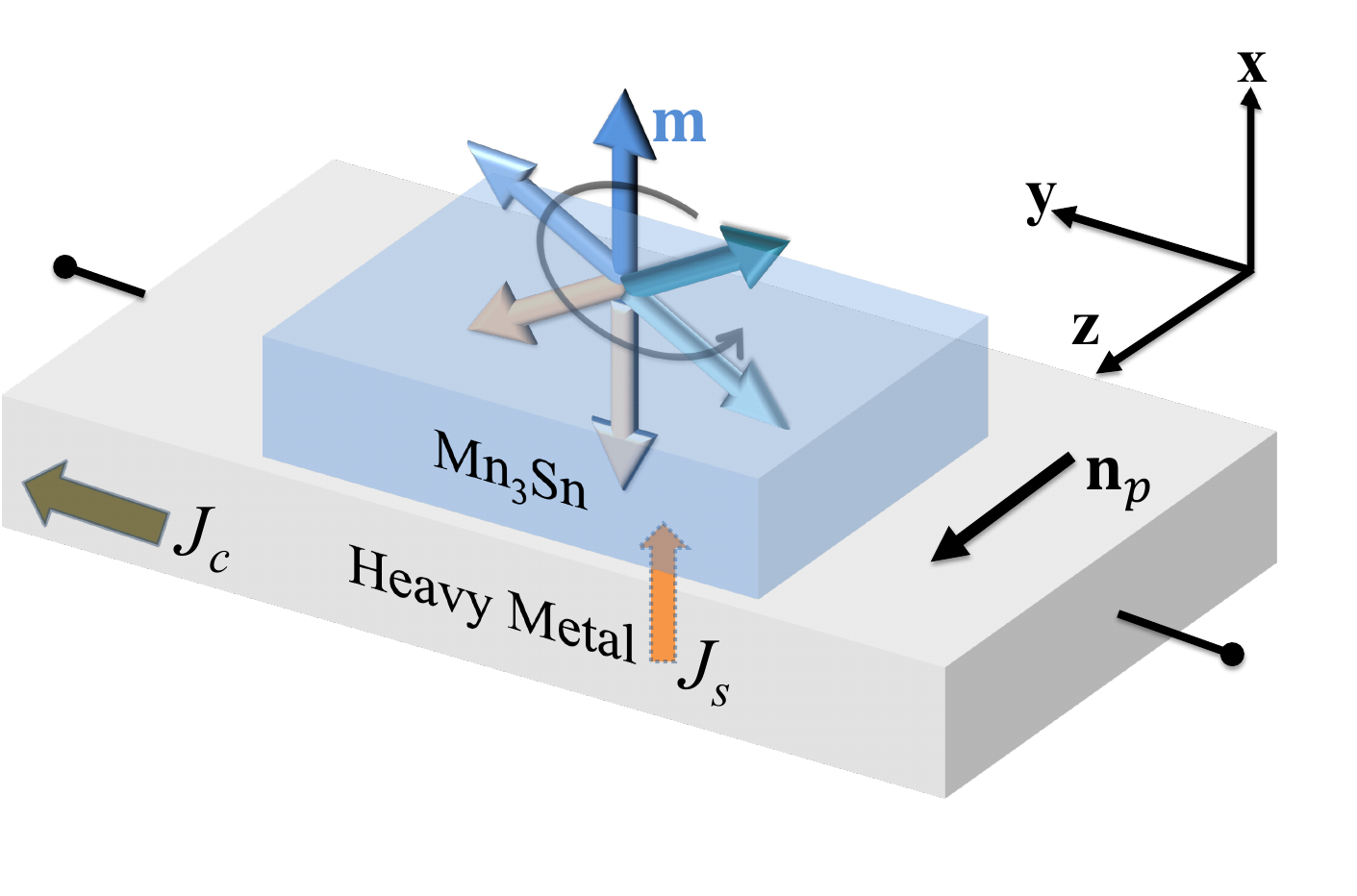}
  \caption{Device setup for manipulating the magnetic state in Mn$_3$Sn. Spin-orbit torque, which is generated due to the spin-Hall effect in the heavy metal, is utilized to manipulate the state of the thin layer of Mn$_3$Sn.
  $\vb{m}$ is the small but nonzero net magnetization in Mn$_3$Sn.
  $J_c$ and $J_s$ are the charge current density and spin current density, respectively.
  } 
  \label{fig:device}
\end{figure} 
\section{SOT-driven dynamics}\label{sec:SOT}
\vspace{-5pt}
The setup used for analyzing the dynamics of the antiferromagnet when subject to SOT is shown in Fig.~\ref{fig:device}. The spin current generated due to charge-to-spin conversion in the heavy metal is polarized along ${\bf{n}}_p$, which coincides with z-axis, per our convention. The Kagome lattice of Mn$_3$Sn is formed in the x-y plane while the z-axis coincides with $[0001]$ direction, as shown in Fig.~\ref{fig:crystal}(b). 
This setup resembles the experimental setup from Refs.~\citen{takeuchi2021chiral, higo2022perpendicular}, where an MgO $(110)[001]$ substrate leads to the selective growth of Mn$_3$Sn in $[0001]$ direction.
The effect of SOT on the AFM order is maximum in this setup since the sublattice vectors predominantly lie in the x-y plane with small z-component while the spin polarization is perpendicular to the x-y plane.~\citep{takeuchi2021chiral}
Other methods of spin injection, such as a spin-polarized electric current from a proximal ferromagnet, could also be used. 
As mentioned previously, our analysis of the SOT dynamics will be carried out in the single-domain limit.  

\begin{figure*}[ht!]
  \centering
  \includegraphics[width = \textwidth, clip = true, trim = 0mm 0mm 0mm 0mm]{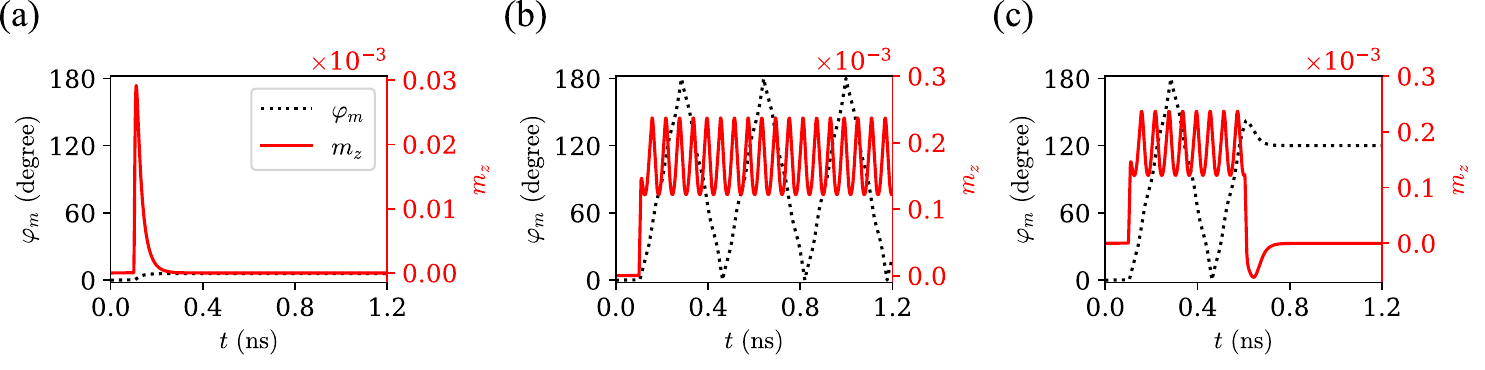}
  \caption{Response of order parameter $\vb{m}$ to different spin currents.
  (a) Non-equilibrium stationary steady-state solution for $J_S = 0.1~\mathrm{MA/cm^2}$. This current is below the threshold current.
  (b) Oscillatory steady-state dynamics for DC current $J_S = 0.5~\mathrm{MA/cm^2}$. 
  (c) Switching dynamics for current pulse of magnitude $J_S = 0.5~\mathrm{MA/cm^2}$ and duration $t_\mathrm{pw} = 500~\mathrm{ps}$. For both (b) and (c) $J_s > J_s^\mathrm{th}$. 
  Here, the thickness of the AFM layer, $d_a = 4~\mathrm{nm}$ and $J_s^\mathrm{th} \approx 1.7~\mathrm{MA/cm^2}$.} 
  \label{fig:sot}
\end{figure*} 
For each sublattice of Mn$_3$Sn, the magnetization dynamics is governed by the classical 
Landau-Lifshitz-Gilbert (LLG) equation, which is a statement of the conservation of angular momentum. The LLG equations for the sub-lattices are coupled via the exchange interaction. 
For sublattice $i$, the LLG equation is given as~\citep{mayergoyz2009nonlinear}
\begin{equation}\label{eq:sLLGS}
    \begin{split}
       \dot{\vb{m}}_i &= - \gamma \mu_{0}\qty(\vb{m}_i \cp \vb{H}_i^\mathrm{eff}) + \alpha \qty(\vb{m}_i \cp \dot{\vb{m}}_i) \\  
        & - \frac{\hbar}{2e}\frac{\gamma J_{s}}{M_{s} d_{a}}\vb{m}_i \cp \qty(\vb{m}_i \cp \vb{n}_{p}),
    \end{split}
\end{equation}
where $\dot{\vb{m}}_{i} = \pdv{\vb{m}_i}{t}$, $t$ is time in seconds, $\vb{H}_{i}^{\mathrm{eff}}$ is the effective magnetic field experienced by $\vb{m}_i$, $\alpha$ is the Gilbert damping parameter for Mn$_3$Sn, $J_s$
is the input spin current density with spin polarization along $\vb{n}_{p} = \vb{z}$, and $d_{a}$ is the thickness of the AFM layer. Other parameters in this equation, viz. $\hbar = 1.054561 \times 10^{-34}~\mathrm{J.s}$, $\mu_{0} = 4\pi \times 10^{-7}~\mathrm{N/A^2}$, $e = 1.6 \times 10^{-19} ~\mathrm{C}$, and $\gamma = 17.6 \cp 10^{10}~\mathrm{T^{-1} s^{-1}}$ are the reduced Planck's constant, the permeability of free space, the elementary charge of an electron, and the gyromagnetic ratio, respectively.
The spin current density depends on the input charge current density, $J_c$, and the spin-Hall angle of the heavy metal (HM), $\theta_\mathrm{SH}$, as $J_s = \theta_\mathrm{SH} J_c$. The spin-Hall angle is associated with the efficiency of the SOT effect.
A recent experiment~\citep{tsai2020electrical} on switching dynamics in Mn$_3$Sn estimated a large (small) and negative (positive) $\theta_\mathrm{SH}$ in W (Pt).
Subsequent experiments~\cite{pal2022setting, higo2022perpendicular} have also preferred W as the heavy metal in their SOT switching experiments; therefore, in this work we consider the HM to be W. Its properties are reported in Table~\ref{tab:mat_params1}.

The effective magnetic field for sublattice $i$ can be obtained by using Eq.~(\ref{eq:energy_density}) as
\begin{flalign}\label{eq:m_field}
    \begin{split}
        \vb{H}_{i}^{\mathrm{eff}}  &= -\frac{1}{\mu_{0} M_{s}}\pdv{F}{\vb{m}_{i}} = -\frac{J_E}{\mu_{0} M_{s}} \qty(\vb{m}_{j} + \vb{m}_k)  \\
        &+\frac{D \vb{z} \cp \qty(\vb{m}_{j} - \vb{m}_{k})}{\mu_{0} M_{s}} + \frac{2K_{e}}{\mu_{0} M_{s}} \qty(\vb{m}_{i} \vdot \vb{u}_{e, i}) \vb{u}_{e, i} + \vb{H}_{a},
    \end{split}
\end{flalign}
where $(i, j, k) = (1, 2, 3), (2, 3, 1),$ or $(3, 1, 2)$, respectively. 
For all the numerical results presented in this work, Eqs.~(\ref{eq:sLLGS}) and~(\ref{eq:m_field}) are solved simultaneously with $\varphi_m^i = 0$ as the initial state. 
The results would be equally applicable if any of the other five equilibrium states were considered as the initial state. 
The dynamic instability caused in the order parameters depends on the magnitude of the SOT relative to the intrinsic energy scale of the AFM. Numerical simulations of the coupled LLG equations show that when the input spin current density ($J_s$) is below a certain threshold current $(J_s^\mathrm{th})$, the order parameters evolve to a non-equilibrium stationary steady-state in the Kagome plane, as shown in Fig.~\ref{fig:sot}(a). 
However, the z-component of the order parameters in the stationary steady-states is zero. 
On the other hand, if $J_s > J_s^\mathrm{th}$, the order parameters exhibit oscillatory dynamics, as shown in Fig.~\ref{fig:sot}(b). The oscillation frequency can be tuned from the GHz to the THz range, depending on the magnitude of $J_s$ relative to $J_s^\mathrm{th}$. 
Finally, Fig.~\ref{fig:sot}(c) shows that if the current is turned off during the oscillatory dynamics, the system could switch to one of the six equilibrium states, $\varphi_m^\mathrm{eq}$. Different final states can be achieved by tuning the duration or the magnitude of the current pulse.

\subsection{Stationary states and threshold current}
\vspace{-5pt}
In order to analytically evaluate the threshold current and the stationary state solutions, we consider $\vb{m}$ as that given by Eq.~(\ref{eq:avg_m}), since our numerical result in Fig.~\ref{fig:sot}(a) indicates it to be in the Kagome plane. 
Next, we evaluate $\dot{\vb{m}} = \frac{\qty(\dot{\vb{m}}_1 + \dot{\vb{m}}_2 + \dot{\vb{m}}_3)}{3}$ as~\citep{zhang2023current}
\begin{flalign}\label{eq:m_dot1}
    \begin{split}
        \dot{\vb{m}} &= -\frac{\gamma \mu_{0}}{3}\qty(\vb{m}_{1} \cp \vb{H}_{1}^\mathrm{eff} + \vb{m}_{2} \cp \vb{H}_{2}^\mathrm{eff} + \vb{m}_{3} \cp \vb{H}_{3}^\mathrm{eff}) \\
        &+ \frac{\alpha}{3} \qty(\vb{m}_1 \cp \dot{\vb{m}}_1 + \vb{m}_2 \cp \dot{\vb{m}}_2 + \vb{m}_3 \cp \dot{\vb{m}}_3) - \frac{1}{3}\frac{\hbar}{2e}\frac{\gamma J_{s}}{M_{s} d_{a}}\\  
        & \times \qty(\vb{m}_1 \cp \qty(\vb{m}_1 \cp \vb{z}) + \vb{m}_2 \cp \qty(\vb{m}_2 \cp \vb{z}) + \vb{m}_3 \cp \qty(\vb{m}_3 \cp \vb{z})),
    \end{split}
\end{flalign}
and then simplify it to 
\begin{flalign}\label{eq:m_dot2}
    \begin{split}
        \dot{\vb{m}} &= -\frac{\gamma}{3M_s}\pdv{F}{\varphi_m} \vb{z} - \alpha \dot{\varphi}_m \vb{z} + \frac{\hbar}{2e}\frac{\gamma J_{s}}{M_{s} d_{a}} \vb{z},
    \end{split}
\end{flalign}
where we have used $\vb{m}_i \cp \vb{H}_{i}^\mathrm{eff} = \frac{-1}{\mu_0 M_s} \pdv{F}{\phi_i} \vb{z} = \frac{-1}{\mu_0 M_s} \pdv{F}{\varphi_m}\pdv{\varphi_m}{\phi_i} \vb{z} = \frac{1}{3\mu_0 M_s} \pdv{F}{\varphi_m} \vb{z}$.
Since the steady-state z-component of $\vb{m}$ must be zero for currents below the threshold current, the net torque is zero in the z-direction viz., $\dot{m}_z = 0$. Therefore, we have
\begin{flalign}\label{eq:phi_dot}
    \alpha \dot{\varphi}_m  = -\frac{\gamma}{M_s} \frac{\qty(3J_E + 7\sqrt{3}D)K_e^3}{9 \qty(J_E + \sqrt{3}D)^3} \sin{\qty(6\varphi_m)} + \frac{\hbar}{2e}\frac{\gamma J_{s}}{M_{s} d_{a}}. 
\end{flalign}

\begin{figure}[ht!]
  \centering
  \includegraphics[width = \columnwidth, clip = true, trim = 0mm 0mm 0mm 0mm]{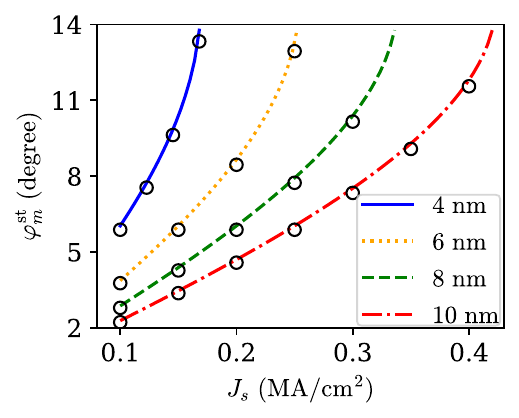}
  \caption{Non-equilibrium stationary steady-states as a function of the applied spin current $J_s$ for different thickness of the Mn$_3$Sn layer. The applied current is below the threshold current in each case. Numerical result from the solution of Eq.~(\ref{eq:sLLGS}) (symbols) fully agree with the analytical results obtained from Eq.~(\ref{eq:stationary}) (lines).} 
  \label{fig:stationary}
\end{figure}

When the spin current is turned on, the energy of the order parameters increases, and they drift away from their equilibrium state. The intrinsic damping of the system, however, dissipates some of the supplied energy.
If the net energy is less than that required to overcome the intrinsic energy barrier, the system settles to a non-equilibrium steady-state, where the antidamping torque due to the spin current is balanced by the torque due to the internal field. Therefore, $\dot{\vb{m}} = 0$, $\dot{\varphi}_m = 0$, and the non-equilibrium stationary steady-states are given as
\begin{equation}\label{eq:stationary}
    \varphi_m^\mathrm{st} = \varphi_m^\mathrm{eq} + \frac{1}{6}\sin^{-1}{\qty(\frac{\frac{\hbar}{2e}\frac{J_{s}}{d_{a}}}{\frac{\qty(3J_E + 7\sqrt{3}D)K_e^3}{9 \qty(J_E + \sqrt{3}D)^3}})},
\end{equation}
where adding $\varphi_m^\mathrm{eq}$ ensures that the above equation is valid irrespective of the initial equilibrium state.
The stationary state of the system as a function of the input spin current density $(J_s < J_s^\mathrm{th})$ is shown in Fig.~\ref{fig:stationary} for different AFM thicknesses. 
It shows excellent agreement between the analytic model of Eq.~(\ref{eq:stationary}) and that obtained from the numerical simulation of Eq.~(\ref{eq:sLLGS}) for all AFM thicknesses examined here.
These results should be widely applicable as long as the thickness of the AFM film is within the single-domain limit.~\citep{higo2022perpendicular, sato2023thermal}
In general, $\varphi_m^\mathrm{st}$ increases with input current since higher current injects more energy and the order parameter moves further away from its equilibrium orientation. 
Finally, the threshold current corresponds to $J_s$ which makes the argument of $\sin^{-1}{(\vdot)}$ in Eq.~(\ref{eq:stationary}) equal to $\pm1$. It is, therefore, given as
\begin{equation}\label{eq:Jth1}
    J_s^\mathrm{th} = d_a\frac{2e}{\hbar}\frac{\qty(3J_E + 7\sqrt{3}D)K_e^3}{9 \qty(J_E + \sqrt{3}D)^3}.
\end{equation}
It corresponds to the minimum current that pushes the order parameter above the energy barrier.
As expected, the strength of $J_s^\mathrm{th}$ depends on the effective energy barrier imposed by the intrinsic energy interactions (Eq.~(\ref{eq:energy_density_phim})).
Similar expressions have also been obtained in case of collinear and non-collinear AFMs with two-fold anisotropy,~\citep{khymyn2017antiferromagnetic, sulymenko2017terahertz, gomonay2015using, puliafito2019micromagnetic, parthasarathy2021precessional, shukla2022spin} however, prior works have not addressed the case of the six-fold anisotropy in materials like Mn$_3$Sn and Mn$_3$Ge.

\subsection{Oscillation dynamics under DC SOT}
\vspace{-5pt}
For currents above the threshold current ($J_s > J_s^\mathrm{th}$), the net energy pumped into the system overcomes the intrinsic barrier imposed by the effective anisotropy of the system. The z-component of the order parameter increases due to the SOT, as shown in Fig.~\ref{fig:sot}(b). 
For the case of the spin polarization perpendicular to the kagome plane, as considered in this work, an equal spin torque acts on all the sublattice vectors along the z-direction.~\citep{higo2022perpendicular, zhang2023current} As a result, their z-components increase in magnitude, which in turn increases the z-component of the average magnetization since $m_z = \qty(m_{1,z} + m_{2,z} + m_{3,z})/3$.
Consequently, an effective exchange field ($\propto J_E m_z$) arises in the z-direction, which leads to oscillation of the order parameters around the z-axis with frequencies in the GHz-THz range.~\citep{yamane2019dynamics, shukla2022spin}  

\begin{figure}[ht!]
  \centering
  \includegraphics[width = \columnwidth, clip = true, trim = 0mm 0mm 0mm 0mm]{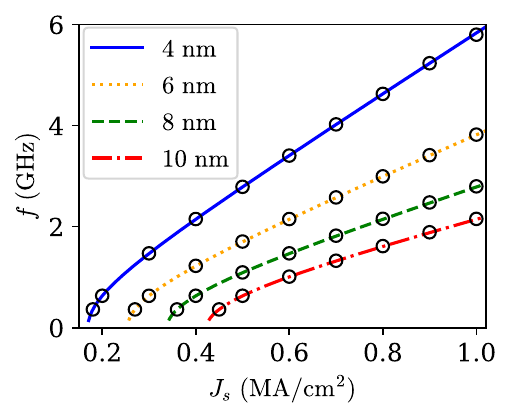}
  \caption{Oscillation frequency as a function of the applied spin current $J_s$ for different thickness of the Mn$_3$Sn layer. The applied current is above the threshold current in each case. Numerical result from the solution of Eq.~(\ref{eq:sLLGS}) (symbols) agree very well with the analytical results obtained from Eq.~(\ref{eq:time_period}) (lines).} 
  \label{fig:frequency}
\end{figure} 
For currents close to the threshold current ($J_s \gtrsim J_s^\mathrm{th}$), $m_z$ is small, as shown in Fig.~\ref{fig:sot}(b). It depends on $\dot{\varphi}_m$ and the exchange constant as $m_z \approx -\frac{\dot{\varphi}_m}{3 \gamma J_E/M_s}$,~\citep{yamane2019dynamics, shukla2022spin} 
while our numerical simulations reveal that $\frac{\dot{m}_z}{\alpha \dot{\varphi}_m} < 0.1$.
Therefore, we can use Eq.~(\ref{eq:phi_dot}) to obtain the time period for $2\pi$ rotation as
\begin{flalign}\label{eq:time_period}
    \begin{split}
        \frac{1}{f} &= \alpha \frac{2e}{\hbar} \frac{M_s d_a}{\gamma J_s}\int_{\varphi_{m, 0}}^{\varphi_{m,0}+2\pi} \frac{d \varphi_m^{'}}{1 - \qty(\frac{J_s^\mathrm{th}}{J_s}) \sin{\qty(6\varphi_m^{'})}} \\
        &=\frac{\alpha \frac{2e}{\hbar} \frac{M_s d_a}{\gamma J_s}}{3\sqrt{1-\qty(\frac{J_s^\mathrm{th}}{J_s})^2}}\tan^{-1}{\qty(\frac{\tan{(3\varphi_m^{'})} - \frac{J_s^\mathrm{th}}{J_s}}{\sqrt{1-\qty(\frac{J_s^\mathrm{th}}{J_s})^2}})}\Bigg|_{0}^{2\pi} \\
        &= \alpha \frac{2e}{\hbar} \frac{M_s d_a}{\gamma J_s} \frac{2\pi}{\sqrt{1 - \qty(\frac{J_s^\mathrm{th}}{J_s})^2}},
    \end{split}
\end{flalign}
where we assumed $\varphi_{m, 0} = 0$.
Figure~\ref{fig:frequency} shows the frequency of oscillation of the order parameter, $f$, as a function of the spin current density for different AFM thicknesses, $d_a$. 
Firstly, for each AFM thickness, $f$ increases with $J_s$ since more energy is pumped into the system.
Secondly, at any current, $f$ decreases with $d_a$ since the strength of spin torque is inversely proportional to $d_a$.
Finally, numerical values of frequency obtained from the solution of Eq.~(\ref{eq:sLLGS}) (symbols) match very well against the results of Eq.~(\ref{eq:time_period}) (lines), thereby showing the efficacy of our models.
The results suggest that close to the threshold current $f$ scales as $\sim J_s^\mathrm{th} \sqrt{\qty(\frac{J_s}{J_s^\mathrm{th}})^2 - 1}$, whereas $f$ increases almost linearly with $J_s$ for $J_s \gg J_s^\mathrm{th}$.
This linear scaling of frequency with the input current has also been suggested in previous works,~\citep{takeuchi2021chiral, shukla2022spin} however, the dependence of the $f$ on the six-fold anisotropy has not been discussed previously.

\subsection{Switching dynamics under pulsed SOT}
\begin{figure}[ht!]
  \centering
  \includegraphics[width = \columnwidth, clip = true, trim = 0mm 0mm 0mm 0mm]{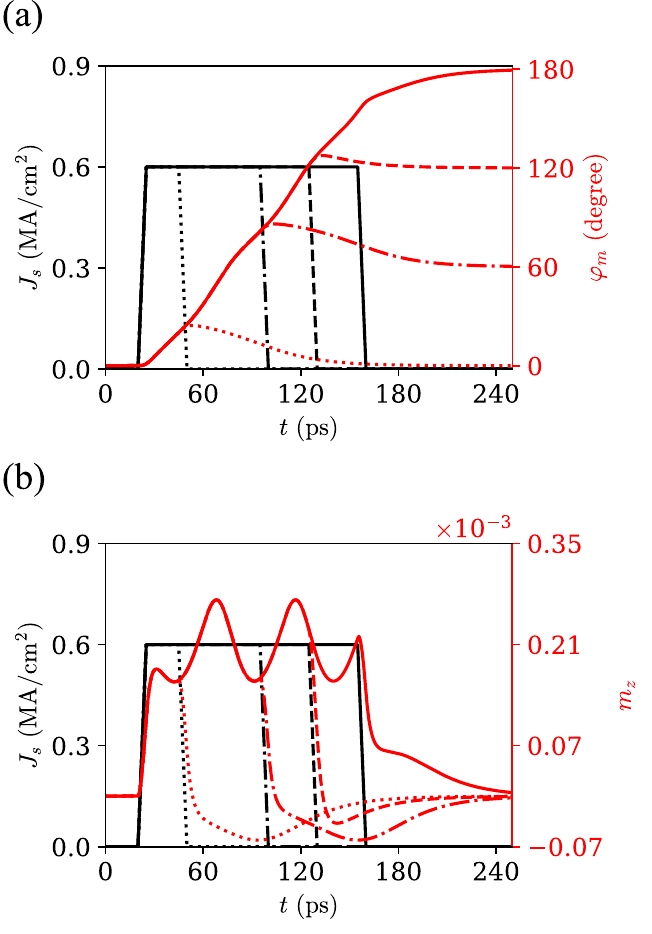}
  \caption{The response of single domain Mn$_3$Sn to a current pulse of fixed amplitude of $0.6~\mathrm{MA/cm^2}$ but different duration. 
  Left axis: Magnitude of current pulses, with rise and fall time of $5~\mathrm{ps}$ each.
  Right axis: (a) The azimuthal angle, $\varphi_m$, as a function of time. 
  (b) The z-component of the average magnetization, $m_z$, as a function of time.
  $\varphi_m$ can be switched from its initial state ($\varphi_m^\mathrm{eq} = 0$) to a different state, if the net input energy (tuned by varying $t_\mathrm{pw}$) overcomes the effective potential barrier. 
  As pulse duration increases $m_z$ stays non-zero for longer period.} 
  \label{fig:pulse_response2}
\end{figure}

\begin{figure}[ht!]
  \centering
  \includegraphics[width = \columnwidth, clip = true, trim = 0mm 0mm 0mm 0mm]{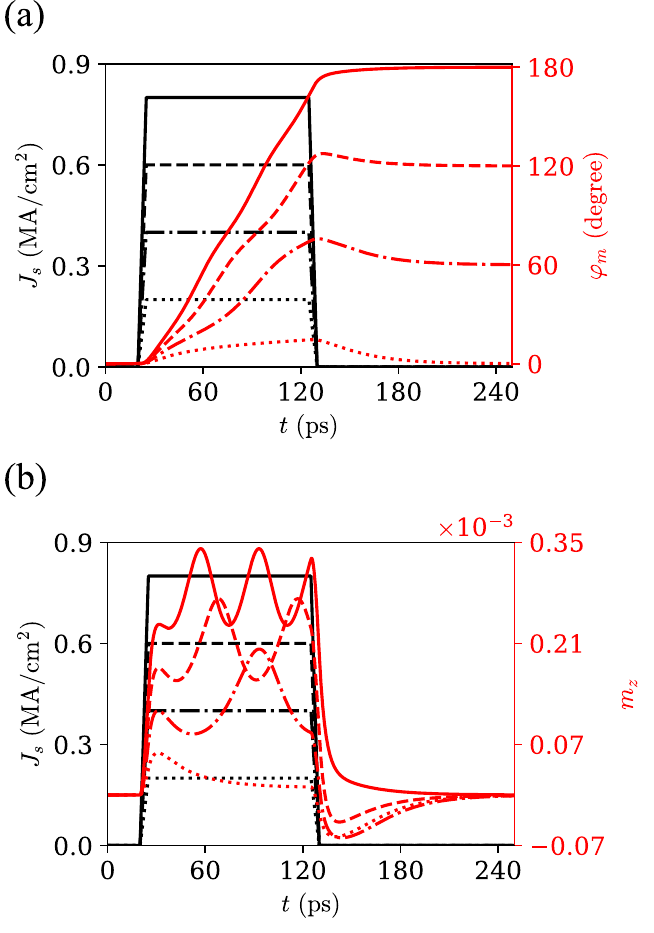}
  \caption{The response of single domain Mn$_3$Sn to a current pulse of fixed duration of $100~\mathrm{ps}$, but different amplitudes. 
  Left axis: Magnitude of current pulses, with rise and fall time of $5~\mathrm{ps}$ each.
  Right axis: (a) The azimuthal angle, $\varphi_m$, as a function of time. 
  (b) The z-component of the average magnetization, $m_z$, as a function of time.
  $\varphi_m$ can be switched from its initial state ($\varphi_m^\mathrm{eq} = 0$) to a different state, if the net input energy (tuned by varying $J_s$) overcomes the effective potential barrier. 
  $m_z$ increases, both in magnitude and frequency, with increase in the pulse amplitude.} 
  \label{fig:pulse_response3}
\end{figure}

\vspace{-5pt}
If the input spin current is turned off during the course of the oscillation, the order parameter loses energy to the intrinsic damping of the AFM. 
As a result, $m_z$ reduces to zero, while $\varphi_m$ settles into the nearest equilibrium state, as shown in Fig.~\ref{fig:sot}(c).
Here, $\pi/2 < \varphi_m < 5\pi/6$ when the current is turned off; therefore, the order parameter settles into $\varphi_m^\mathrm{eq} = 2\pi/3$ equilibrium state.
Comparing Figs.~\ref{fig:sot}(b) and~\ref{fig:sot}(c), one can conclude that it is possible to switch to all six states, if the current pulse is switched off at an appropriate time. 
Indeed the same is shown in Fig.~\ref{fig:pulse_response2}, where different final states are achieved by changing the duration of the current pulse of magnitude $J_s = 0.6~\mathrm{MA/cm^2}$.
Another way to switch between stable states is by varying the magnitude of current pulse for a fixed duration. Figure~\ref{fig:pulse_response3} shows switching of the order parameter from $\varphi_m^\mathrm{eq} = 0$ to $\varphi_m^\mathrm{eq} = \pi/3, 2\pi/3$ and $\pi$ as the magnitude of the input current is increased, while its duration is fixed at 100 ps. 
In Figs.~\ref{fig:pulse_response2} and~\ref{fig:pulse_response3}, we only present switching in one direction, that is from $\varphi_m^\mathrm{eq} = 0$ through $\varphi_m^\mathrm{eq} = \pi$ since $\varphi_m^\mathrm{eq} = 4\pi/3$ and $\varphi_m^\mathrm{eq} = 5\pi/3$ could be achieved by changing the direction of input current.
Finally, $m_z (\propto - \dot{\varphi}_m)$ increases both in magnitude and frequency with $J_s$ (Fig.~\ref{fig:pulse_response3}(b)) since $\dot{\varphi}_m$ increases with current (Eq.~\ref{eq:phi_dot}). 

Increasing either the magnitude or the duration of the current pulse increases the input energy to the order parameter. As a result, it starts a precessional motion till the current pulse is turned off, after which it loses its energy to the intrinsic damping and reaches an equilibrium state.
The final state could, therefore, be considered a function of the net input energy.
For low energy switching operation between two equilibrium states, the net input energy should be just enough to move the order parameter to the top of a barrier preceding the final equilibrium state, thereafter the intrinsic damping of the system assists in the switching process. 
To determine the duration ($t_\mathrm{pw}$) and magnitude ($J_s$) of the current pulse required to switch from $\varphi_m^\mathrm{eq} = 0$ to $\varphi_m^\mathrm{eq} = n\pi/3$ $\qty(n = \{1, 2, 3\})$, we use Eq.~(\ref{eq:phi_dot}).
We find that $t_\mathrm{pw}$ and $J_s$ required to switch to a different state depends on the final state as
\begin{flalign}\label{eq:delay}
    \begin{split}
        J_s t_\mathrm{pw} &= \alpha \frac{2e}{\hbar}\frac{M_s d_a}{\gamma}\frac{\frac{(2n-1) \pi}{2} + \tan^{-1}{\qty(\frac{J_s^\mathrm{th}/J_s}{\sqrt{1-\qty(J_s^\mathrm{th}/J_s)^2}})} }{3\sqrt{1 - \qty(J_s^\mathrm{th}/J_s)^2}}. 
    \end{split}
\end{flalign}

\begin{figure}[ht!]
  \centering
  \includegraphics[width = \columnwidth, clip = true, trim = 0mm 0mm 0mm 0mm]{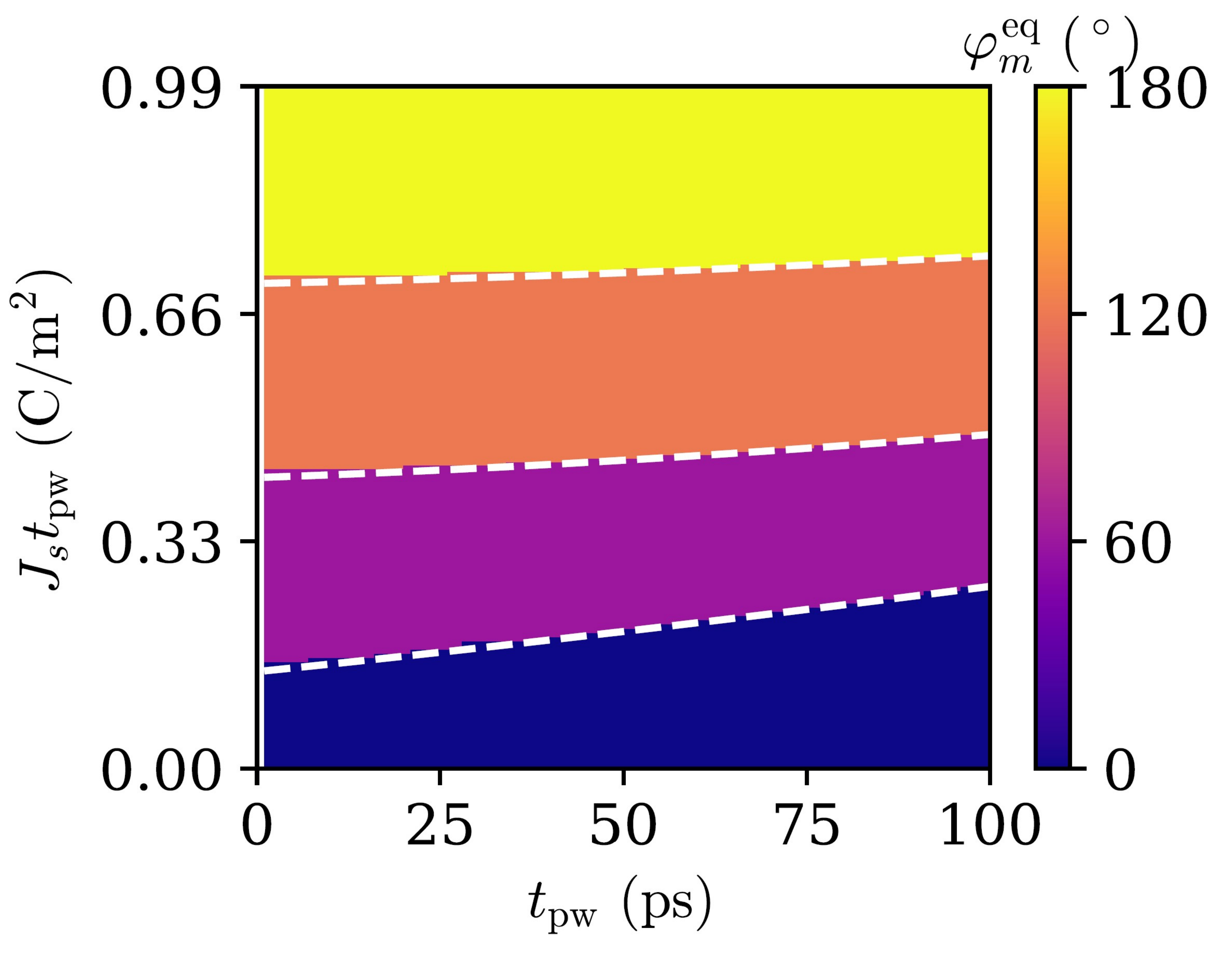}
  \caption{The final state of $\vb{m}$ as a function of the pulse duration, $t_\mathrm{pw}$, and the total injected spin charge density, $J_s t_\mathrm{pw}$. 
  The overlaid dashed white lines represent the analytic bounds of Eq.~(\ref{eq:delay}).
  Here, the thickness of the AFM film is $d_a = 4~\mathrm{nm}$.}
  \label{fig:final_angle}
\end{figure}

This equation suggests that for large currents $(J_s \gg J_s^\mathrm{th}$), the final equilibrium state depends only on the input spin charge density ($J_s t_\mathrm{pw}$) as $J_s t_\mathrm{pw} \sim \alpha \frac{2e}{\hbar}\frac{M_s d_a}{\gamma} \frac{(2n-1) \pi}{6}$. 
Such a large current density above the threshold current density corresponds to the small pulse width regime.
For current magnitude near the threshold current $(J_s \gtrsim J_s^\mathrm{th}$) and large $t_\mathrm{pw}$, the final state scales almost linearly with $t_\mathrm{pw}$ as $J_s^\mathrm{th} t_\mathrm{pw} \sqrt{\qty(\frac{J_s}{J_s^\mathrm{th}})^2 -1 } \sim \alpha \frac{2e}{\hbar}\frac{M_s d_a}{\gamma} \frac{n\pi}{3}$.
To verify these claims, we solved Eq.~(\ref{eq:sLLGS}) for $t_\mathrm{pw}$ in the range $[1, 100]~\mathrm{ps}$ and spin charge density in the range $[0, 1)~\mathrm{C/m^2}$, and evaluated the final states. 
It can be observed from Fig.~\ref{fig:final_angle} that for small $t_\mathrm{pw}$, the final states $(\varphi_m^\mathrm{eq})$ is indeed dependent only on $J_s t_\mathrm{pw}$, more evidently for $\varphi_m^\mathrm{eq} = 2\pi/3$ and $\pi$.
This is because the input current is large for these cases. 
On the other hand, the final states for large pulse width clearly depends on $t_\mathrm{pw}$.
This behavior is more noticeable for $\varphi_m^\mathrm{eq} = \pi/3$ and $\varphi_m^\mathrm{eq} = 2\pi/3$ whereas for $\varphi_m^\mathrm{eq} = \pi$, it would be evident for longer pulse widths. 
The overlaid dashed lines correspond to Eq.~(\ref{eq:delay}) for $n = {1, 2, 3}$ and represent the minimum spin charge density required to switch from $\varphi_m^\mathrm{eq} = 0$ to $\varphi_m^\mathrm{eq} = n\pi/3$ as a function of the pulse width.
The fluctuations of the Hall resistance under the effect of SOT in Ref.~\citen{takeuchi2021chiral} were credited to the aforementioned pulsed dynamics. However, an expression describing the interdependence of $J_s$ and $t_\mathrm{pw}$ required to switch from one known state to another known state, as a function of the material parameters, was missing.

\begin{figure}[ht!]
  \centering
  \includegraphics[width = \columnwidth, clip = true, trim = 0mm 0mm 0mm 0mm]{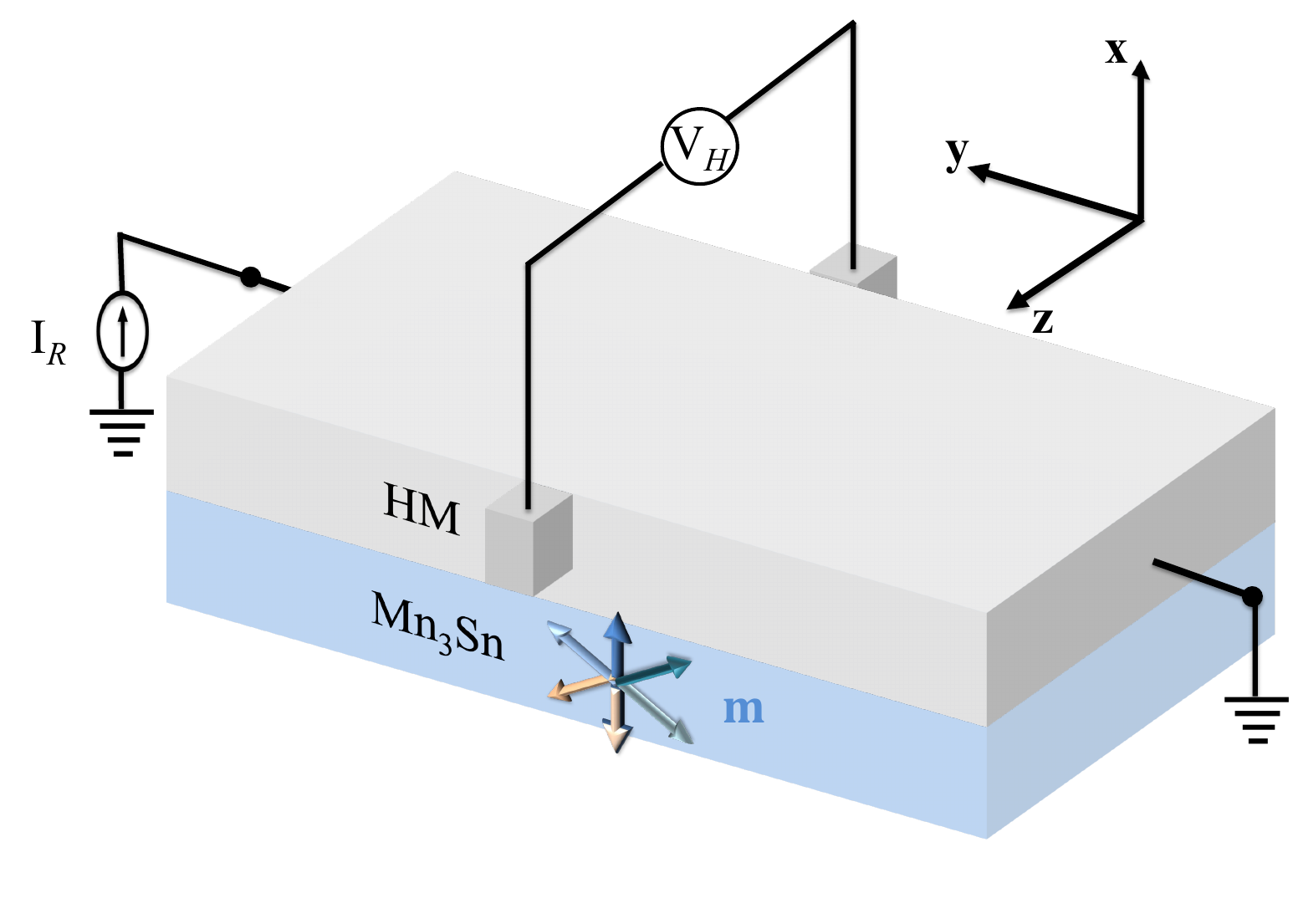}
  \caption{A Hall bar setup to measure the output voltage signal, $V_\mathrm{H}$, generated on the injection of charge read current, $I_\mathrm{R}$. This current is lower than the threshold current to induce dynamics, $I_c^\mathrm{th}$.
  The arrows show different orientations of $\vb{m}$.} 
  \label{fig:Hallbar}
\end{figure}

\section{Readout of magnetic state}\label{sec:detect}
\vspace{-5pt}
\subsection{Anomalous Hall Effect}\label{subsec:AHE}
\vspace{-5pt}
Non-collinear AFMs, such as Mn$_3$Sn, exhibit non-zero AHE due to the non-zero Berry curvature in the momentum space when certain symmetries are absent.~\citep{kubler2014non}
In a Hall bar setup, a read current $(I_R)$ lower than the threshold charge current ($I_c^\mathrm{th}$) to induce dynamics, is applied to a bilayer of the heavy metal and Mn$_3$Sn, as shown in Fig.~\ref{fig:Hallbar}.
A voltage $(V_H)$ transverse to the current flow is measured to detect the magnetization state of Mn$_3$Sn. Since this voltage depends on the Hall resistance of the Mn$_3$Sn layer, $R_H^\mathrm{Mn_3Sn}$, it is given as~\citep{tsai2021large, higo2022perpendicular}
\begin{flalign}\label{eq:AHE}
    \begin{split}
        V_H = I_R \qty(1 + \frac{\rho_\mathrm{Mn_3Sn}}{\rho_\mathrm{HM}} \frac{d_\mathrm{HM}}{d_\mathrm{Mn_3Sn}})^{-2} R_H^\mathrm{Mn_3Sn},
    \end{split}
\end{flalign}
where $\rho_\mathrm{Mn_3Sn} (\rho_\mathrm{HM})$ and $d_\mathrm{Mn_3Sn} (d_\mathrm{HM})$ are the electrical resistivity and thickness of the Mn$_3$Sn (HM) layer, respectively. The Hall resistance of Mn$_3$Sn depends on the Hall resistivity $\rho_H^\mathrm{Mn_3Sn}$, and is given as $R_H^\mathrm{Mn_3Sn} = \frac{\rho_H^\mathrm{Mn_3Sn}}{d_\mathrm{Mn_3Sn}}$. 
In the above equation, the magnitude of current leakage into the AFM layer is evaluated assuming a parallel resistance combination of the heavy metal and the AFM layers, as $I_R^\mathrm{Mn_3Sn} = I_R\qty(1 + \frac{\rho_\mathrm{Mn_3Sn}}{\rho_\mathrm{HM}} \frac{d_\mathrm{HM}}{d_\mathrm{Mn_3Sn}})^{-1}$.
On the other hand, the transverse Hall voltage can be obtained as $V_H = V_H^\mathrm{Mn_3Sn}\qty(1 + \frac{\rho_\mathrm{Mn_3Sn}}{\rho_\mathrm{HM}} \frac{d_\mathrm{HM}}{d_\mathrm{Mn_3Sn}})^{-1}$, where the bilayer is assumed as two resistances connected in series.

To calculate the Hall voltage in our device setup, we consider $\rho_\mathrm{HM} = 43.8~\mathrm{\mu \Omega \vdot cm}$ and $d_\mathrm{HM} = 7~\mathrm{nm}$, corresponding to that of W (listed in Table~\ref{tab:mat_params1}), while the other parameters are taken as $\rho^\mathrm{Mn_3Sn} = 330~\mathrm{\mu \Omega \vdot cm}$,~\citep{takeuchi2021chiral} $d_\mathrm{Mn_3Sn} = 4~\mathrm{nm}$, and $|\rho_H^\mathrm{Mn_3Sn}| = 3~\mathrm{\mu \Omega \vdot cm}$.~\citep{sung2018magnetic}
If a read current of $I_R = 50~\mathrm{\mu A}$ is applied to the bilayer, a Hall voltage of $V_H = 1.86~\mathrm{\mu V}$ is generated in the HM layer. 
As indicated in Eq.~(\ref{eq:AHE}), a larger read current would increase~\citep{tsai2021large} the detected Hall voltage. Alternatively, a larger Hall resistivity of Mn$_3$Sn would be desired to generate larger read voltages. 
Since the read and write terminals in this device setup are same, this setup is more suitable for detecting the switching dynamics.  

\begin{figure}[ht!]
  \centering
  \includegraphics[width = \columnwidth, clip = true, trim = 0mm 0mm 0mm 0mm]{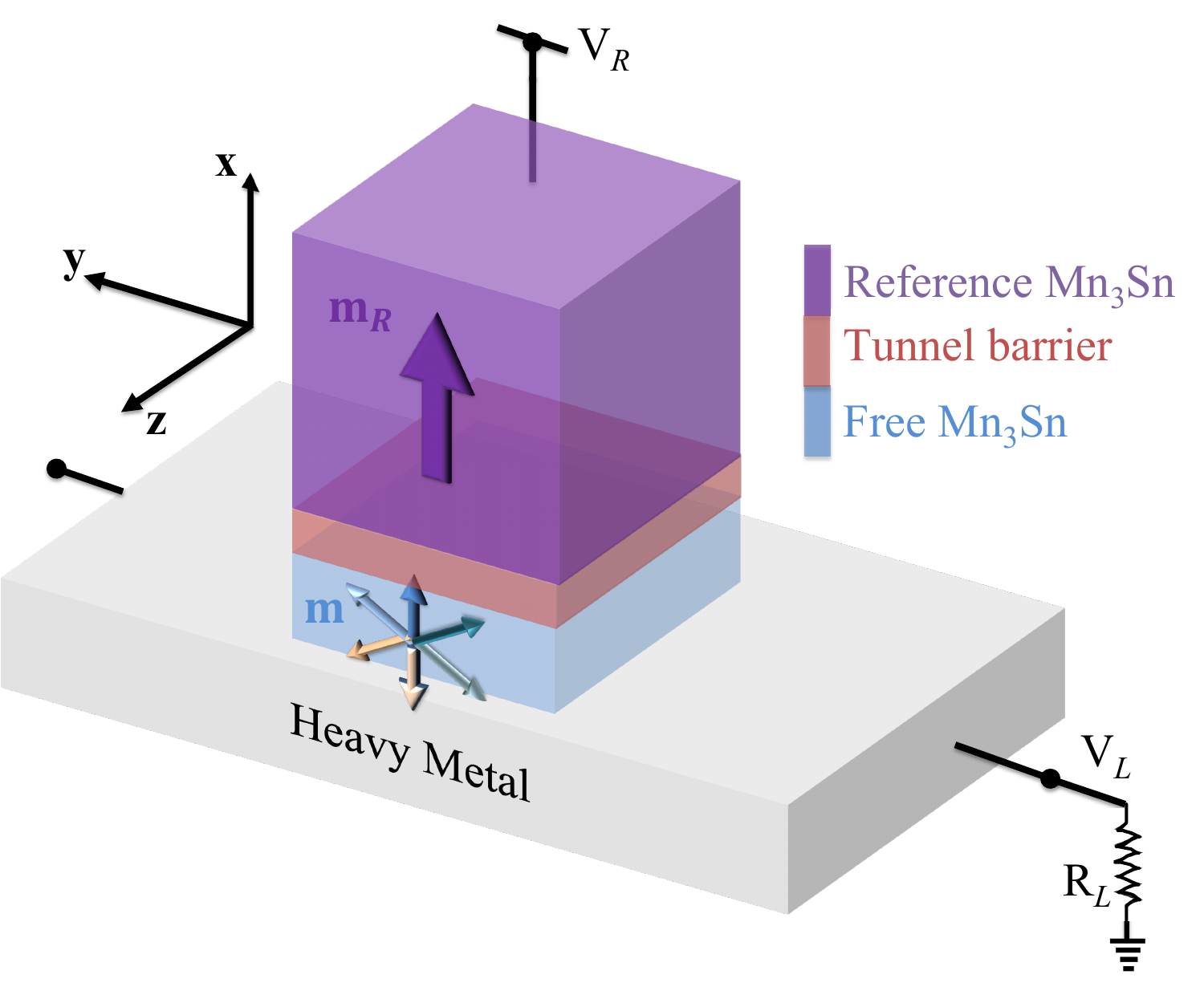}
  \caption{All antiferromagnetic tunnel junction atop the heavy metal layer. Both the reference and free layers are composed of Mn$_3$Sn.
  The arrows in free Mn$_3$Sn show the six equilibrium orientations of $\vb{m}$ while that in reference Mn$_3$Sn shows the fixed orientation of its order parameter, $\vb{m}_R$.
  A small read voltage, $V_R$, is applied and a voltage, $V_L$, is detected across the resistive load, $R_L$.
  $V_L$ depends on the orientation of $\vb{m}$ with respect to $\vb{m}_R$.} 
  \label{fig:aftj}
\end{figure}
\subsection{Tunneling Magnetoresistance}
\vspace{-5pt}
An all antiferromagnetic tunnel junction built using two Mn$_3$Sn electrodes has recently been shown to exhibit a finite TMR at room temperature.~\citep{chen2023octupole}
It was experimentally demonstrated that a TMR value of about $2\%$ at room temperature was possible for thicker tunnel barrier (MgO) and large resistance-area (RA) product. The high (low) resistance state corresponds to an angle of $\pi (0)$ between the magnetic octupoles of the two Mn$_3$Sn layers.
The electronic bands in Mn$_3$Sn show momentum-dependent spin splitting of the Fermi surface due to the broken time reversal symmetry.~\citep{dong2022tunneling, chen2023octupole}
As a result, a spin-polarized current is generated.
The relative orientation of the magnetic octupoles in the two AFM layers controls the magnitude of the spin-polarized current, and hence the magnetoresistance.~\citep{dong2022tunneling} 
A simple representation of an all-Mn$_3$Sn tunnel junction is shown in Fig.~\ref{fig:aftj}. Here, a thick layer of Mn$_3$Sn, with fixed orientation $\vb{m}_R$, is used as the reference layer, while the order parameter $\vb{m}$ of the thinner layer is modulated using the device setup of Fig.~\ref{fig:device}. The different orientations of $\vb{m}$ with respect to $\vb{m}_R$ is measured as the read voltage, $V_L$, across a resistive lead, $R_L$, resulting in a finite read voltage, $V_R$. 

Theoretical calculations have revealed that for vacuum as a tunnel barrier, RA product increased from approximately $0.05~\mathrm{\Omega \vdot \mu m^2}$ in the parallel configuration to $0.2~\mathrm{\Omega \vdot \mu m^2}$ in the antiparallel configuration.~\citep{dong2022tunneling} 
For the states corresponding to $\varphi_m^\mathrm{eq} = \pi/3$ and $\varphi_m^\mathrm{eq} = 2\pi/3$, the RA product is evaluated to be approximately $0.07~\mathrm{\Omega \vdot \mu m^2}$ and $0.11~\mathrm{\Omega \vdot \mu m^2}$, respectively.~\citep{dong2022tunneling}
Assuming the same value of RA product in our device setup, we estimate the voltage drop across $R_L = 50~\mathrm{\Omega}$ as $V_L = 9.1~\mathrm{mV}$ for the parallel configuration, while $V_L = 7.1~\mathrm{mV}$ for the antiparallel configuration of the free and fixed layers of the tunnel junction.
On the other hand, $V_L = 8.8~\mathrm{mV}$ and $V_L = 8.2~\mathrm{mV}$ for $\varphi_m^\mathrm{eq} = \pi/3$ and $\varphi_m^\mathrm{eq} = 2\pi/3$, respectively.
Here, we considered $V_R = 10~\mathrm{mV}$ while the area of the tunnel barrier was assumed to be $\mathcal{A} = 100 \times 100~\mathrm{nm^2}$. 
This setup can be used for detecting both the oscillation and switching dynamics.

\vspace{-5pt}
\section{Consideration of thermal effects }\label{sec:temp}
\vspace{-5pt}
When Mn$_3$Sn is driven by large DC SOT to excite oscillatory dynamics in the 100's of GHz range, it is likely that Joule heating in the device structure would impose practical limits on the operating conditions. For example, our simulations show that we need approximately $17~\mathrm{MA/cm^2}$ spin current to generate oscillations of the order parameter at 100 GHz in a $4~\mathrm{nm}$ thick film of Mn$_3$Sn. Due to the spin Hall angle of heavy metal being less than unity, the required charge current could be 2-20$\times$ higher than the spin current. The large charge current requirement will increase the temperature of the device structure, which, if significant, could alter the properties of the magnetic material, or in extreme cases change the magnetic order completely. 
In addition to the magnitude of the input current pulse, the Joule heating in the device also depends on the electrical properties and the size of the heavy metal that the charge current passes through.
The temperature rise of the device due to the generated heat depends on the ambient temperature and the thermal properties of the conducting (magnetic and non-magnetic) media and the substrate. 
Finally, in the pulsed response case, the time-domain temperature profile is expected to be sensitive to the duration of the applied current and thus the peak temperature in this case could be markedly different from that achieved when DC charge currents are used in the device.

Recent experimental investigation of field-assisted current-driven switching dynamics in Mn$_3$Sn attribute the switching process to Joule heating.~\citep{pal2022setting, krishnaswamy2022time} 
They show that despite the spin diffusion length in Mn$_3$Sn being extremely small (roughly $1~\mathrm{nm}$), the measured AHE signals showed switching behavior for films of thicknesses ranging from $30~\mathrm{nm}$ to $100~\mathrm{nm}$.
Subsequently, this switching dynamics was explained as a demagnetization of the AFM order in Mn$_3$Sn due to Joule heating over a time scale of 10's of nanosecond (ns), followed by a remagnetization of the AFM order due to cooling in the presence of SOT.  
We must, however, point out that the SOT device setups in these experiments were different from our setup of Fig.~\ref{fig:device}.
Unlike our case, the spin polarization associated with the SOT was in the Kagome plane, which increases the current requirement.~\citep{takeuchi2021chiral, higo2022perpendicular}
Nevertheless, an accurate investigation of the current-driven magnetization dynamics in our device setup, requires a careful consideration of the temperature rise due to Joule heating. 

To investigate the temperature rise in our setup, we assume a one-dimensional heat flow into the substrate (MgO), as presented in Ref.~\citen{coffie2003characterizing}.
Consequently, the junction temperature is strongly depended on the properties of the substrate such as its thermal conductivity $\kappa_\mathrm{sub}= 40~\mathrm{W/(m \vdot K)}$, thickness $d_\mathrm{sub}$, mass density $\rho_\mathrm{sub} = 3580~\mathrm{kg/m^3}$, and specific heat capacity $C_\mathrm{sub} = 930~\mathrm{J/(kg \vdot K)}$.
These values are taken from Ref.~\citen{meinert2018electrical}.
In our model, we assume that the HM and Mn$_3$Sn are at the same temperature and that the temperature is spatially invariant in the bilayer. 
The area of the heat source is set by the dimensions, i.e., the length and the width, of the bilayer. 
The rise in the junction temperature, $\Delta T_j (t)$, is then calculated as~\citep{coffie2003characterizing} 
\begin{flalign}\label{eq:DeltaTj}
    \begin{split}
        &\Delta T_j (t) = \Delta T_M\left(\qty(1 - \frac{8}{\pi^2} \sum_{k = \mathrm{odd}}\frac{\exp(-k^2 t/\tau)}{k^2}) \right. \\
        &\left. - \theta(t - t_\mathrm{pw}) \qty(1 - \frac{8}{\pi^2} \sum_{k = \mathrm{odd}}\frac{\exp(-k^2 (t - t_\mathrm{pw})/\tau)}{k^2})\right),
    \end{split}
\end{flalign}
where
\begin{equation}
    \Delta T_M = \frac{d_\mathrm{sub} J_c^2}{\kappa_\mathrm{sub}}\frac{\qty(\rho_\mathrm{Mn_3Sn}d_\mathrm{Mn_3Sn} +   \qty(\frac{\rho_\mathrm{Mn_3Sn}}{\rho_\mathrm{HM}}\frac{d_\mathrm{HM}}{d_\mathrm{Mn_3Sn}})^2\rho_\mathrm{HM}d_\mathrm{HM})}{\qty(1 + \frac{\rho_\mathrm{Mn_3Sn}}{\rho_\mathrm{HM}} \frac{d_\mathrm{HM}}{d_\mathrm{Mn_3Sn}})^2}
\end{equation}
is the maximum possible rise in the junction temperature for an input charge current density, $J_c$, and includes the current leakage into the Mn$_3$Sn layer (as described in Section~\ref{subsec:AHE}). 
The rate of the heat flow into the substrate is determined by the thermal rate constant, $\tau = \frac{4 d_\mathrm{sub}^2 \rho_\mathrm{sub} C_\mathrm{sub}}{\pi^2 \kappa_\mathrm{sub}}$.
Finally, $\theta(\cdot)$ in Eq.~(\ref{eq:DeltaTj}) is the Heaviside step function.

\begin{figure}[ht!]
  \centering
  \includegraphics[width = \columnwidth, clip = true, trim = 0mm 0mm 0mm 0mm]{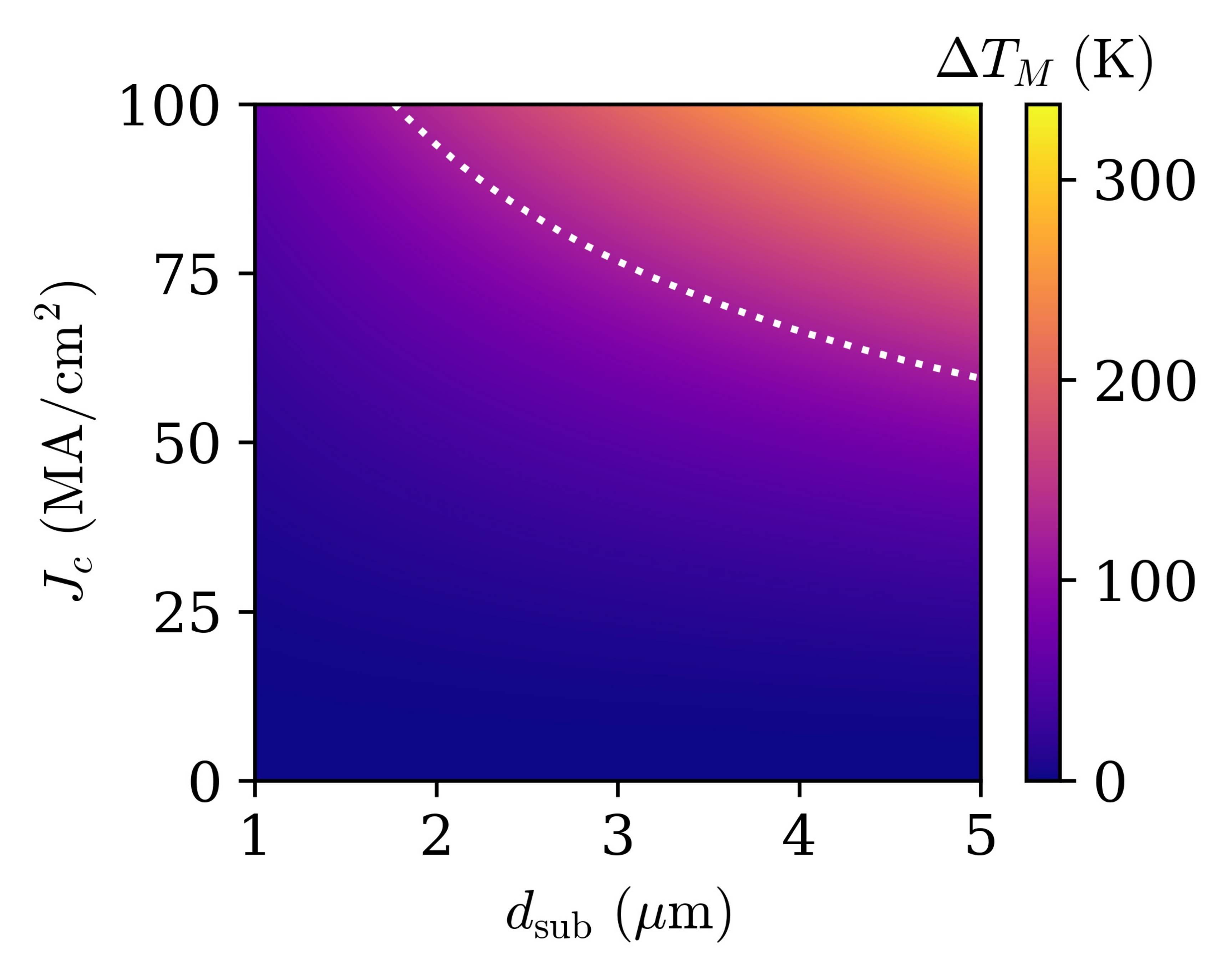}
  \caption{Maximum possible rise in the junction temperature, $\Delta T_M$, for different substrate thickness, $d_\mathrm{sub}$, and input charge current density, $J_c.$ 
   $\Delta T_\mathrm{M}$ increases with both $d_\mathrm{sub}$ and $J_c$, as expected.
   The dotted white line corresponds to $\Delta T_M = 120~\mathrm{K}$, which is the maximum temperature rise before the antiferromagnetic order in Mn$_3$Sn changes.} 
  \label{fig:temp_heatmap}
\end{figure} 

Figure~\ref{fig:temp_heatmap} shows the maximum rise in the junction temperature for different substrate thicknesses and DC input charge current densities.
As expected, $\Delta T_M$ increases with both $d_\mathrm{sub}$ and $J_c$.
An increase in $J_c$, increases the heat generated in the metallic layers, whereas an increase in $d_\mathrm{sub}$, increases the thermal resistance in the direction of heat flow.
Assuming the heat sink temperature to be $300~\mathrm{K}$, $\Delta T_M \gtrsim 120~\mathrm{K}$ would change the antiferromagnetic order since the N\`eel temperature for Mn$_3$Sn is approximately $420~\mathrm{K}$.
If we consider $d_\mathrm{sub} = 2~\mathrm{\mu m}$, then $\Delta T_M < 120~\mathrm{K}$ for $J_c < 9.4 \times 10^7~\mathrm{A/cm^2}$. This limits the use of a $4~\mathrm{nm}$ thick Mn$_3$Sn film as a signal generator to about $33~\mathrm{GHz}$.
On the other hand, for $d_\mathrm{sub} = 5~\mathrm{\mu m}$, the charge current density is limited to $J_c < 6 \times 10^7~\mathrm{A/cm^2}$, which limits the frequency of the signal source to about $21~\mathrm{GHz}$.
The white dotted line overlaid on Fig.\ref{fig:temp_heatmap} represents this upper limit for the charge current for different substrate thicknesses.
A thinner substrate, or the metallic heat sources with smaller area of cross-section could help lower the maximum temperature rise.

\begin{figure}[ht!]
  \centering
  \includegraphics[width = \columnwidth, clip = true, trim = 0mm 0mm 0mm 0mm]{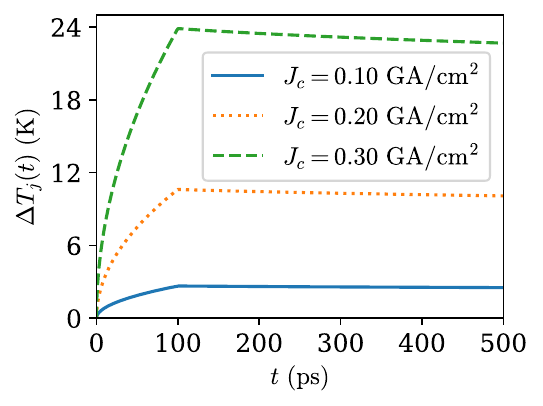}
  \caption{Temperature rise of HM-Mn$_3$Sn bilayer due to Joule heating associated with the charge current pulse of different amplitudes, as indicated in the legend, and 100 ps duration. 
  For $J_c = 3 \times 10^8~\mathrm{A/cm^2}$, the temperature rise of the junction is approximately $24~\mathrm{K}$. On the other hand, for $J_c = 1 \times 10^8~\mathrm{A/cm^2}$ and $J_c = 2 \times 10^8~\mathrm{A/cm^2}$ the maximum temperature rise is approximately $2.7~\mathrm{K}$ and $10.6~\mathrm{K}$, respectively. Once the input current pulse is turned off, the temperature begins to recover to the ambient temperature; however, the recovery is slow due to the large time constant of heat flow associated with the substrate.} 
  \label{fig:temp}
\end{figure} 
The time to reach $\Delta T_M$, on the other hand, depends on $\tau$. For a constant input current, $\Delta T_j (t)$ reaches to approximately $95\%$ of $\Delta T_M$ in $t = 3\tau$.
If we consider $d_\mathrm{sub} = 1~\mathrm{\mu m}$, then $\tau = 33.7~\mathrm{ns}$, and the time to reach $95\%$ of the steady-state temperature is approximately $100~\mathrm{ns}$.
Consequently, input currents larger than $J_c = 1.3 \times 10^8~\mathrm{A/cm^2}$ could be safely applied, if the duration of the pulse is of the order of few 10's of $\mathrm{ps}$.
Figure~\ref{fig:temp} shows the non quasi-static response of the rise in junction temperature, $\Delta T_j (t)$, as a function of time for current pulses of duration $t_\mathrm{pw} = 100~\mathrm{ps}$.
It can be observed that even for charge current density as large as $J_c = 3 \times 10^8~\mathrm{A/cm^2}$, the temperature rise is less than $24~\mathrm{K}$.
Therefore, such large currents could be used during the switching operation, if required.
Similar behavior is expected for other values of $t_\mathrm{pw} \ll \tau$.
A drawback of the high value of $\tau$, due to large $d_\mathrm{sub}$, is the slow removal of heat from the system, resulting in a slow decrease of temperature, as shown in Fig.~\ref{fig:temp}.  
A thinner substrate or a substrate with lower thermal capacitance is preferred to reduce the time for heat to flow from the source to the sink, decreasing $\tau$. 

The aforementioned analysis does not consider the temperature dependence of the material parameters (electrical and thermal) of the AFM, the HM, or the substrate.
In general, the material properties of the AFM, such as the uniaxial anisotropy constant, could decrease with an increase in temperature due to Joule heating.~\citep{han2023coherent}
Therefore, if the input current is slowly increased from zero to a higher value, the stationary steady-state solutions could be different from those given by Eq.~(\ref{eq:stationary}) while
the onset of dynamics could occur at an input current lower than that predicted by Eq.~(\ref{eq:Jth1}).
The oscillation frequency at an input current (Eq.~(\ref{eq:time_period})), on the other hand, could increase owing to the lower threshold current as well as Joule heating induced lower saturation magnetization. Consequently, the order parameter could be switched at a faster rate using a current pulse of smaller duration since $J_s t_\mathrm{pw}$ (Eq.~(\ref{eq:delay})) would decrease.
However, an exact value of the threshold current, stationary state, oscillation frequency, or minimum pulsewidth required to switch between two states as a function of the input current would depend on the relative change in the antiferromagnetic material parameters with temperature.

Non-zero temperature leads to random fluctuation of the AFM spins, and as a result the order parameter would exhibit a distribution around the equilibrium states in the six degenerate energy wells.
When acted upon by external spin current the order parameter would exhibit stochastic dynamics---each stationary steady-state below the threshold current would show a distribution around a mean value, the oscillation frequency would exhibit a non-zero linewidth, and $J_s t_\mathrm{pw}$ would shown thermal noise-induced randomness.
The effect of thermal fluctuations would be more prominent below and near threshold current while its effect on the dynamics would be weak for large values of input currents.
Similar to ferromagnets, thermal noise in AFMs could be modeled as three random Gaussian fields, one for each sublattice, with zero mean and standard deviation equal to one.~\citep{go2022noncollinear}
In our modeling approach, this field would be added to Eq.~(\ref{eq:m_field}), and the total field would consist of those due to exchange interaction, DM interaction, uniaxial anisotropy, external field, and thermal noise. 
The goal of our work, however, was to understand the deterministic dynamics in monodomain Mn$_3$Sn, and therefore the effect of thermal noise is not considered here.

\section{Conclusion and Outlook}
In this work, we investigated the SOT-driven oscillation and switching dynamics in six-fold magnetic anisotropy degenerate single-domain Mn$_3$Sn thin films. The spin polarization associated with the SOT was assumed perpendicular to the Kagome plane. Numerical simulations of the SOT-driven dynamics in the DC regime reveal that the frequency of oscillation of the order parameter in Mn$_3$Sn could be tuned  
from 100's of MHz to 100's of GHz by increasing the input spin current density. Analytic models of frequency versus input spin current density presented here for the first time show an excellent agreement with the numerical results, obtained by solving the classical coupled LLG equations of motion for the sublattice vectors of Mn$_3$Sn. 
In addition to the oscillatory dynamics, switching of the order parameter between the six energy minima could also be accomplished by utilizing pulsed input spin current. In this case, the switching of the order parameter can be controlled by tweaking the amplitude and the pulse width of the input spin current. To provide a physical insight into the physics and functionality of Mn$_3$Sn in the pulsed-SOT regime,
we developed analytic models of the final state that the order parameter settles into as a function of the spin charge density and revealed excellent match with numerical results. 
Motivated by recent experimental advances in electrical detection of the magnetic state of Mn$_3$Sn, we discussed AHE and TMR as possible schemes to detect the state of the AFM in a microelectronics-compatible manner. We showed that AHE voltage signals could be in the range of $\sim \mu$V, while the TMR voltage signals could be around $7-9~\mathrm{mV}$ with a further possibility of increasing this voltage range for higher RA product values. 
Finally, we evaluated the temperature rise as a function of the input current pulse. Our results showed that the maximum temperature rise for MgO substrate of thickness $d_\mathrm{sub} = 1~\mathrm{\mu m}$ is less than $24~\mathrm{K}$ for current pulses with $t_\mathrm{pw} \leq 100~\mathrm{ps}$ and $J_c \leq 3 \times 10^8~\mathrm{A/cm^2}$. On the other hand, in the DC mode, the maximum temperature rise is found to be $120~\mathrm{K}$ for $J_c = 1.3 \times 10^8~\mathrm{A/cm^2}$, thereby limiting the application of Mn$_3$Sn as a coherent signal generator for frequencies up to $45~\mathrm{GHz}$.
In this work, we considered thin films with thicknesses less than $10~\mathrm{nm}$. 
Typically, such films show two-fold degeneracy, instead of the six-fold, owing to epitaxial strain.
This will be addressed in our future work.

\section*{Supplementary Material}
See supplementary material for the perturbative analysis of the ground state.

\begin{acknowledgments}
This research was primarily supported by the NSF through the University of Illinois at Urbana-Champaign Materials Research Science and Engineering Center DMR-1720633.
\end{acknowledgments}

\section*{Data Availability Statement}
The data that support the findings of this study are available from the corresponding author upon reasonable request.

\nocite{*}
\bibliography{APLMat_Mn3Sn}

\end{document}